\documentclass[singlecolumn]{article}

\usepackage{arxiv}

\usepackage[utf8]{inputenc} 
\usepackage[T1]{fontenc}    
\usepackage{hyperref}       
\usepackage{url}            
\usepackage{booktabs}       
\usepackage{amsfonts}       
\usepackage{nicefrac}       
\usepackage{microtype}      
\usepackage{graphicx}
\usepackage[sort]{natbib}
\setcitestyle{comma,numbers,open={[},close={]}}
\usepackage{doi}

\title{A physics-augmented neural network framework for finite strain incompressible viscoelasticity}

\author{
	Karl A. Kalina\\
	Chair of Computational and\\
	Experimental Solid Mechanics\\
	TU Dresden,
	01062 Dresden, Germany \\
	\And
	J\"{o}rg Brummund\\
	Chair of Computational and\\
	Experimental Solid Mechanics\\
	TU Dresden,
	01062 Dresden, Germany \\
	\And
	Markus K\"{a}stner\thanks{Corresponding author, email: \texttt{markus.kaestner@tu-dresden.de}.} \\
	Chair of Computational and\\
	Experimental Solid Mechanics\\
	TU Dresden, 
	01062 Dresden, Germany \\
}

\renewcommand{\shorttitle}{A physics-augmented neural network framework for finite strain incompressible viscoelasticity}

\hypersetup{
pdftitle={Preprint_KalinaEtAl_2025},
pdfsubject={},
pdfauthor={Kalina},
pdfkeywords={First keyword, Second keyword, More},
}

\usepackage{amssymb}
\usepackage{amsthm}
\usepackage{amsmath}
\usepackage{siunitx}
\usepackage{mathtools}		
\mathtoolsset{centercolon}	
\usepackage{newtxmath}      
\usepackage{lineno}
\usepackage{bbm}
\usepackage[dvipsnames]{xcolor}
\usepackage{multirow}
\usepackage[normalem]{ulem}
\usepackage{algorithm2e}
\RestyleAlgo{ruled}
\usepackage{siunitx}

\newcommand{\R}{\mathbb R} 
\newcommand{\N}{\mathbb N} 
\newcommand{\Sym}{\mathscr{S\! y \! m}} 
\newcommand{\Devi}{\mathscr{D\! e \! v}} 
\newcommand{\Ln}{\mathcal L} 
\newcommand{\SO}{\mathscr{S\!O}(3)} 
\newcommand{\Othree}{\mathscr{O}(3)} 
\newcommand{\GLp}{\mathscr{G\!L}^+(3)} 
\newcommand{\SL}{\mathscr {S\!L}(3)} 
\newcommand{\incs}{\mathscr{I\!n\!c}} 
\newcommand{\cali}{\mathscr{C\!a\!l}} 

\newcommand{\B}{\mathcal{B}} 

\newcommand{\I}{\boldsymbol{\mathcal I}} 

\DeclareMathOperator{\tr}{tr}
\DeclareMathOperator{\cof}{cof}
\DeclareMathOperator{\sym}{sym}

\DeclareMathOperator{\diag}{diag}
\DeclareMathOperator{\dev}{dev}
\DeclareSIUnit[number-unit-product = \,]{\promille}{\textperthousand}

\newcommand{\dx}{\mathrm d} 
\newcommand{\ve}[1]{\boldsymbol{#1}} 
\newcommand{\te}[1]{\boldsymbol {#1}} 
\newcommand{\tttte}[1]{\mathbb #1} 
\newcommand{\ttttes}[1]{\mathbbm #1} 
\newcommand{\one}{\textit{\textbf{1}}}
\newcommand{\zero}{\textit{\textbf{0}}}

\newcommand{\vect}[1]{\uline{\mathbf{#1}}} 
\newcommand{\matr}[1]{\uuline{\mathbb{#1}}} 

\newcommand{\bte}[1]{\bar{\boldsymbol #1}} 
\newcommand{\bteT}[1]{\bar{\boldsymbol #1}{}^T} 
\newcommand{\bCi}{\bar{\boldsymbol C}{}^\text{i}} 
\newcommand{\bCe}{\bar{\boldsymbol C}{}^\text{e}} 
\newcommand{\mEl}{{}_\xi\!} 
\newcommand{\mElt}[1]{{}_\xi^{#1}\!} 
\newcommand{\mEltl}[1]{{}_{\;\;\;\xi}^{#1}\!} 
\newcommand{\mEltll}[1]{{}_{\;\;\xi}^{#1}\!} 

\newcommand{\mte}[1]{\boldsymbol{\mathcal #1}} 

\newcommand{\diffp}[2]{\frac{\partial #1}{\partial #2}} 
\newcommand{\difft}[2]{\frac{\mathrm{d} #1}{\mathrm{d} #2}} 

\newcommand{\nablaX}{\nabla_{\!\!{\ve X}}}

\makeatletter
\newcommand*\bigcdot{\mathpalette\bigcdot@{.5}}
\newcommand*\bigcdot@[2]{\mathbin{\vcenter{\hbox{\scalebox{#2}{$\m@th#1\bullet$}}}}}
\makeatother

\theoremstyle{definition} 
\newtheorem{rmk}{Remark}

\newtheorem{theorem}{Theorem}
\newtheorem{lemma}[theorem]{Lemma}
\newtheorem{proposition}[theorem]{Proposition}

\usepackage[font=small]{caption}

\begin{document}

\maketitle

\begin{abstract}
 We propose a physics-augmented neural network (PANN) framework for finite strain incompressible viscoelasticity within the generalized standard materials theory. The formulation is based on the multiplicative decomposition of the deformation gradient and enforces unimodularity of the inelastic deformation part throughout the evolution.
Invariant-based representations of the free energy and the dual dissipation potential by monotonic and fully input-convex neural networks ensure thermodynamic consistency, objectivity, and material symmetry by construction.
The evolution of the internal variables during training is handled by solving the evolution equations using an implicit exponential time integrator. In addition, a trainable gate layer combined with $\ell_p$ regularization automatically identifies the required number of internal variables during training.
The PANN is calibrated with synthetic and experimental data, showing excellent agreement for a wide range of deformation rates and different load paths. We also show that the proposed model achieves excellent interpolation as well as plausible and accurate extrapolation behaviors.
In addition, we demonstrate consistency of the PANN with linear viscoelasticity by linearization of the full model.
\end{abstract}


\section{Introduction}
\label{sec:int}

Constitutive models are fundamental to solid mechanics as they provide a mathematical framework for describing the behavior of various materials such as metals or elastomers. Over the past century, extensive research has been carried out to define the physical and mathematical principles that these models should satisfy \cite{Silhavy1997,Haupt2000, Holzapfel2000}. This has led to the development of numerous so-called \emph{classical constitutive models}. 
However, when applied to soft materials that show a highly nonlinear and inelastic behavior, these models are often not accurate enough and may need to be modified if applied to new experimental data. 
To overcome these limitations, \emph{machine learning} approaches -- in particular \emph{neural networks (NNs)} -- have emerged as powerful tools for constitutive modeling \cite{Dornheim2023,Fuhg2024b}. These data-driven methods offer  flexibility to capture complex material responses and automate the process of constitutive modeling.

\subsection{Constitutive modeling with neural networks}

In their seminal work from the early 1990s, Ghaboussi~et~al.~\cite{Ghaboussi1991} were the first to apply neural networks -- specifically, \emph{feedforward neural networks (FNNs)} -- to model hysteresis under both uniaxial and multiaxial stress conditions. To capture the history-dependent nature of material behavior, the FNN was supplied with input data from multiple previous time steps. Although neural network-based constitutive modeling saw some initial interest in the 1990s, it was not actively pursued for quite some time afterward.
However, with the recent surge in machine learning popularity and improvements in computational efficiency, a variety of data-driven techniques\footnote{Besides NNs, other machine learning methods have been explored for constitutive modeling, such as Gaussian process regression \cite{Frankel2020,Ellmer2024}. Additionally, splines have been used to define elastic energy \cite{Wiesheier2024}. Approaches like sparse or symbolic regression have enabled automated discovery of constitutive models \cite{Flaschel2021,Flaschel2023,Meyer2023a,Abdusalamov2023}, allowing algorithms to identify models from a broad candidate space.}
have rapidly gained momentum in the field of mechanics, as reviewed in \cite{Bock2019,Liu2021,Dornheim2023,Fuhg2024b}.

A crucial development in NN-based constitutive modeling and scientific machine learning in general is the incorporation of fundamental physical concepts, which is referred to as \emph{physics-informed} \cite{Raissi2019,Henkes2022,Bastek2023}, \emph{mechanics-informed} \cite{Asad2022}, \emph{physics-augmented} \cite{Klein2024,Linden2023}, \emph{physics-based} \cite{Aldakheel2025,Baktheer2025}, \emph{physics-constrained} \cite{Kalina2023}, or \emph{thermodynamics-based} \cite{Masi2021}. 	
This can be achieved in two ways: either strongly, as in the case of network architectures tailored to the problem \cite{Kalina2022a,Linka2021}, or weakly, as in the case of problem-specific loss functions for training, see \cite{Rosenkranz2023,Weber2023,Geiger2025}. As shown in \cite{Linden2023,Masi2021,Fuhg2023,Masi2024}, these models enable the use of sparse training data and a significant improvement in the model's extrapolation capability.
In the following, we will give a short overview on NN-based constitutive modeling for \emph{elasticity}, \emph{elasto-plasticity} and \emph{viscoelasticity}.

There are numerous works that model elasticity with NNs, whereby the most common approach is to use architectures with the \emph{hyperelastic potential} as output and \emph{invariants} as inputs, e.g., \cite{Linden2023,Klein2021,Kalina2022a,Thakolkaran2022,Linka2021,Fuhg2022b,Tac2024a,Bahmani2024,Benady2024,Peirlinck2024}.
Thereby, a special training technique labeled as \emph{Sobolev training} \cite{Czarnecki2017,Vlassis2020} allows direct calibration of the NN using stress and strain tuples. In particular, the loss function involves the gradient of the energy w.r.t. the deformation.
In addition, \emph{polyconvex} NNs are used in several works \cite{Klein2021,Tac2022a,Chen2022,Tac2024a,Bahmani2024,Jadoon2025a,Dammass2025b}, which improves the extrapolation capability \cite{Linden2023,Kalina2024} and guarantees \emph{rank-one convexity} and thus \emph{ellipticity} \cite{Ebbing2010,Schroder2010}.    
The most widely spread technique to incorporate this is the application of \emph{fully input convex neural networks (FICNNs)} introduced by Amos~et~al.~\cite{Amos2017}. 
It should be noted that polyconvex models based on invariants may be too restrictive for the precise fitting of some data sets \cite{Kalina2024,Klein2026}. However, in the special case of isotropy, polyconvex models based on FICNNs and principal stretches \cite{Vijayakumaran2025} or signed singular values \cite{Geuken2025a} are even more flexible than models based on the invariants $I_1,I_2,I_3$ and thus offer an alternative.

The literature also contains a large number of NN models for modeling \emph{inelastic behavior} that are based on a rigorous physical framework. Many of these approaches use the concept of internal variables.
\emph{Elasto-plastic} models for small strains are presented in \cite{Masi2021,Vlassis2021,Malik2021}, whereby thermodynamic consistency in \cite{Masi2021} is only weakly fulfilled by a loss term. Furthermore, knowledge of the internal variables is required for training. Although these can be obtained from homogenization simulations using autoencoders \cite{Masi2022}, the application of approaches that require internal variables to be prescribed for training is not practical in real experiments.
In \cite{Meyer2023a,Fuhg2023}, elasto-plastic NN models that are thermodynamically consistent by construction are formulated for small deformations. Furthermore, training is performed without prescribed internal variables by solving the evolution equations in each optimization step.
Elasto-plastic models extended for \emph{finite deformations} are presented in \cite{Boes2024,Jadoon2025}.

An important NN-based approach to model viscoelastic behavior is presented by Huang~et~al.~\cite{Huang2022}. The model is embedded in the \emph{generalized standard materials (GSMs)} framework, i.e., thermodynamic consistency is ensured by the use of a dissipation potential that is convex w.r.t. the internal variables as well as normalized and stationary for rates of zero, or alternatively by a dual dissipation potential with equivalent properties, but which depends on the thermodynamic forces.
Several approaches based on a similar modeling strategy can be found, e.g., \cite{Rosenkranz2023,Rosenkranz2024,Flaschel2025}. In contrast to \cite{Huang2022}, however, it is not necessary to prescribe internal variables during training. Only the number needs to be specified. An approach based on the multiplicative split of the deformation gradient and using neural ordinary differential equations (NODEs) is considered in \cite{Tac2023}.
Likewise, models using the multiplicative split can be found in connection with a co-rotational formulation in \cite{Holthusen2024,Holthusen2024a,Holthusen2026}. In addition, a dual dissipation potential approach that also ensures thermodynamic consistency but is based on a less restrictive convexity requirement is introduced in \cite{Holthusen2026}. Therein, the potential only needs to be convex, stationary and normalized in a modified invariant set and not w.r.t. the thermodynamic forces itself.
Another finite strain NN model for viscoelasticity based on GSMs is presented in \cite{Asad2023}, whereby the multiplicative split of the deformation gradient is not assumed. Finally, \cite{Abdolazizi2023a} presents a finite strain model that builds on the generalized Prony series, and \cite{Califano2026} introduces a deep rheological element that models the viscosity via NNs.

\subsection{Objectives and contributions of this work}

As discussed in the literature overview given above, numerous approaches to model \emph{finite strain viscoelasticity} exist that combine modern machine learning methods with a reasonable physical basis. 
Thereby, NN models that use invariants and are embedded into the GSM framework seem most promising as they allow to enforce material symmetry as well as thermodynamic consistency by construction. 
Finite strain models that use the multiplicative decomposition and are based on general NN approaches have so far only been discussed in the work by Tac~et~al.~\cite{Tac2023}, which is based on NODEs, and the very recent approach by Holthusen~et~al.~\cite{Holthusen2025}, which was developed almost simultaneously with our work. The last paper introduces a compressible anisotropic NN model with a weakened non-convex dual potential and uses RNNs as auxiliary networks to provide internal variables during training.

Thus, to the best of the authors' knowledge, there are no works that provide a \emph{finite strain viscoelastic model} that is based on the \emph{multiplicative decomposition}, is \emph{incompressible}, enforces \emph{unimodularity of the inelastic deformation} during evolution and uses \emph{general NN ansatzes} for the potentials in combination with an algorithmic implementation that allows for the application to \emph{multiaxial deformation states}, training with \emph{implicit time discretization} schemes as well as an \emph{automatic determination} of the \emph{number of internal variables} based on $\ell_p$ regularization.
We therefore present such a model in this article, which follows the idea of physics-augmented neural networks (PANNs). To this end, we introduce a rigorous theory for finite strain incompressible viscoelasticity that is embedded into the GSM framework and uses complete invariant sets. In addition, we show a linearization for the case of small strains and provide an \emph{exponential map} time integrator valid for multiaxial states. Based on these concepts, PANNs for the description of the free energy and the dual dissipation potential are introduced. To enable robust training, we introduce several stabilization techniques for the constrained optimization problem to be solved.The model is calibrated with synthetic as well as real experimental data.

The organization of the remaining paper is as follows: In Sect.~\ref{sec:framework}, the underlying GSM framework is presented. After this, PANNs for the description of the potentials as well as a training method are introduced in Sect.~\ref{sec:pann}. The developed approach is exemplarily applied to several examples in Sect.~\ref{sec:examples}. After a discussion of the results, the paper is closed by concluding remarks and an outlook to necessary future work in Sect.~\ref{sec:conc}.   

\paragraph{Notation}
Within this work, tensors of rank one and two are given by boldface italic letters, i.e., $\ve A, \ve B \in \Ln_1$ or $\te C, \te D \in \Ln_2$, where $\Ln_n$ denotes the space of tensors with rank $n\in \N$ with $\N$ being the set of natural numbers without zero.
Tensors with rank four are marked by blackboard symbols, i.e., $\tttte A \in \Ln_4$.
Single and double contractions of two tensors are given by $\ve C \cdot \te D = C_{kl} D_{li} \ve e_k \otimes \ve e_i$ and $\te C:\te D=C_{kl}D_{kl}$, respectively. Therein, $\ve e_k\in \Ln_1$ and $\otimes$ denote a Cartesian basis vector and the dyadic product, where the Einstein summation convention is used. 
Transpose and inverse of a 2nd order tensor $\te C$ are given by $\te C^T$ and $\te C^{-1}$, respectively.
Additionally, $\tr \te C$, $\det \te C$, $\cof \te C := \det(\te C) \te C^{-T}$, $\sym \te C$ and $\dev \te C:=\te C - 1/3 \tr (\te C) \one$ are used to indicate trace, determinant, cofactor as well as symmetric and deviatoric part, respectively.
The sets \mbox{$\Sym:=\left\{\te A \in \Ln_2 \, |\, \te A = \te A^T\right\}$} and \mbox{$\Sym_4:=\left\{\tttte A \in \Ln_4 \, |\, A_{ijkl} = A_{jikl} = A_{ijlk} = A_{klij} \right\}$} denote the spaces of  symmetric 2nd order tensors and 4th order tensors with major and minor symmetry.
Furthermore, the orthogonal group and special orthogonal group  are given by $\Othree:=\left\{\te A \in \Ln_2\,|\,\te A^T \cdot \te A = \one\right\}$ and $\SO:=\left\{\te A \in \Ln_2\,|\,\te A^T \cdot \te A = \one,\,\det \te A = 1\right\}$, respectively, while $\GLp:=\left\{\te A \in \Ln_2\,|\,\det \te A > 0\right\}$ is the set of invertible 2nd order tensors with positive determinant, $\SL:=\left\{\te A \in \Ln_2\,|\,\det \te A = 1\right\}$ the special linear group and $\Devi:=\left\{\te A \in \Ln_2 \, | \, \tr \te A = 0 \right\}$ the set of deviatoric 2nd order tensors.
Thereby, $\one:=\delta_{ij}\ve e_i \otimes \ve e_j\in \Ln_2$ is the 2nd order identity tensor, where $\delta_{ij}$ denotes the Kronecker delta. Similarly, the 4th order identity tensor with major symmetry as well as major and minor symmetry are defined as $(\ttttes 1)_{ijkl} := \delta_{ik}\delta_{jl}$ and $(\ttttes 1^\text{s})_{ijkl} := 1/2(\delta_{ik}\delta_{jl} + \delta_{il}\delta_{jk})$, respectively.
Norms of rank one and two tensors or matrices are given by $|\ve A| := \sqrt{A_iA_i}$ and $\|\te C\| := \sqrt{C_{ij}C_{ij}}$, respectively.

For reasons of readability, the arguments of functions are usually omitted within this work. However, potentials are given with their arguments to show the dependencies, except when derivatives are written. Furthermore, in the following the symbol of a function is identical with the symbol of the function value itself.

\section{Finite strain incompressible viscoelasticity modeling framework}
\label{sec:framework}

In this section, we introduce kinematics and stress measures common in finite strain continuum theory. Afterwards, a framework for the modeling of \emph{incompressible finite strain viscoelasticity} based on the concept of \emph{GSMs} is presented.
In addition, the model is transferred to the linear theory with small strains using Taylor series expansion.
Finally, we introduce appropriate time integration schemes.

\subsection{Kinematics and stress measures}
\label{sec:kinematics_stress}

\paragraph{Kinematics}
Let us consider the motion of a material body with reference configuration \mbox{$\mathcal{B}_0 \subset \R^3$} at time $t_0 \in \R_{\ge 0}$  and current configurations \mbox{$\mathcal{B}_t \subset \R^3$} at times \mbox{$t\in \mathcal T:=\{\tau\in \R \,|\,\tau \ge t_0\}$}. To describe the body's motion, we introduce smooth bijective mappings $\ve \varphi_t: \mathcal{B}_0 \to \mathcal{B}_t$, mapping material points $\ve X\in\mathcal{B}_0$ to $\ve x_t=\ve \varphi_t(\ve X) \in \mathcal{B}_t$. In order to enable the calculation of derivatives w.r.t. time later on, we represent the mappings $\ve \varphi_t(\ve X)$ as a function of space and time in what follows, i.e., $\ve \varphi(\ve X,t)$ \cite[Sect.~2.2]{Silhavy1997}. With that, the displacement $\ve u \in \Ln_1$ of each material point is given by $\ve u(\ve X, t) := \ve \varphi(\ve X, t) - \ve X$
and the velocity is defined as $\ve v := \dot{\ve u}$, where $\dot{(\bullet)}$ is the material time derivative.

As additional kinematic quantities, the deformation gradient $\te F := (\nablaX \ve \varphi)^T \in \GLp$ and the Jacobi determinant \mbox{$J:=\det \te F \in \R_{>0}$} are defined. Using the Flory split \cite{Flory1961}, we introduce the isochoric part of the deformation gradient $\bte F:=J^{-1/3} \te F\in\SL$ with $\det \bte F = 1$. Based on these quantities, we introduce the symmetric and positive definite right Cauchy-Green deformation tensor \mbox{$\te C:=\te F^T \cdot \te F \in \Sym \cap \GLp$} and its isochoric part \mbox{$\bte C:=\bteT F \cdot \bte F \in \Sym \cap \SL$} as well as the Green-Lagrange strain tensor
$\te E:= 1/2 (\te C - \one) \in \Sym$ as kinematic quantities which are invariant to rigid body motions. Finally, we define the velocity gradient $\te l :=(\nabla \ve v)^T\in\Ln_2$ and the deformation rate $\te d:=\sym(\te l)\in\Sym$.

\paragraph{Stress measures}
Within finite strain continuum mechanics, several stress measures can be defined. Here, we make use of the \emph{Cauchy stress} tensor $\te \sigma \in \Sym$, which is also known as true stress, as well as the \emph{1st and 2nd Piola-Kirchhoff stress} tensors $\te P \in \Ln_2$ and $\te T \in \Sym$. The latter two stress measures are linked to the Cauchy stress by the pull-back operations $\te P:= J \te \sigma \cdot \te F^{-T}$ and $\te T:= J \te F^{-1}\cdot \te \sigma \cdot \te F^{-T}$, respectively.

For more details on basic principles in continuum solid mechanics the reader is referred to the textbooks of {\v S}ilhav{\'y}~\cite{Silhavy1997}, Haupt~\cite{Haupt2000} or Holzapfel~\cite{Holzapfel2000}.

\subsection{Modeling of viscoelasticity with generalized standard materials}

\begin{figure}
	\centering
	\includegraphics{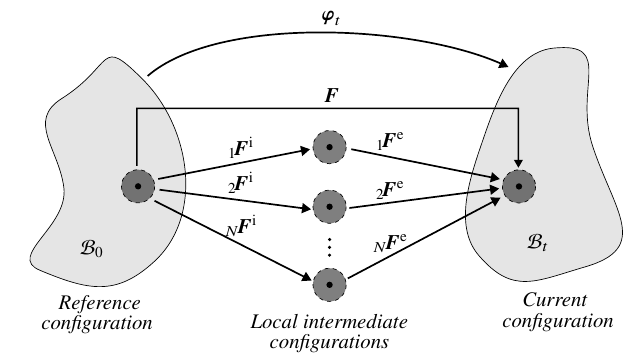}
	\caption{Visualization of fictitious intermediate configurations  $\mEl\B^\text{i}$ implied by the multiplicative decompositions $\te F := \mEl\te F^\text{e} \cdot \mEl\te F^\text{i}$ in finite strain viscoelasticity modeling. Figure inspired by \cite{Yamakawa2021}.}
	\label{fig:multF}
\end{figure}

Now we introduce a general framework for the modeling of \emph{isotropic incompressible finite strain viscoelasticity},
where we build up on a \emph{generalized Maxwell-type} model with $N\in\N$ Maxwell elements, that is illustrated in Fig.~\ref{fig:maxwell}.

\subsubsection{Multiplicative decomposition}
We begin by discussing the kinematics of deformation-like internal variables.	Our description assumes $N$ \emph{multiplicative decompositions} of the deformation gradient and its Jacobi determinant
\begin{align}
	\te F := \mEl\te F^\text{e} \cdot \mEl\te F^\text{i}
	\text{ and } J = \mEl J^\text{e} \mEl J^\text{i}
	\;, \; \xi \in\{1,2,\ldots,N\} \label{eq:mult}
\end{align}
into elastic parts $\mEl\te F^\text{e}\in\GLp$ and inelastic parts $\mEl\te F^\text{i}\in \GLp$ related to the dissipation \cite{LeTallec1993,Reese1998,Bergstrom1998a,Kumar2016,Rambausek2022,Ciambella2024,Holthusen2024}. Thereby, $\mEl J^\text{e} := \det \mEl \te F^\text{e}$ and $\mEl J^\text{i} := \det \mEl \te F^\text{i}$. Thus, $\mEl\te F^\text{e}$ and $\mEl\te F^\text{i}$ take on the role of internal variables.

\begin{rmk}
	\label{rmk:mult}	
	The split into elastic and inelastic parts according to Eq.~\eqref{eq:mult} can be interpreted by introducing fictitious inelastic intermediate configurations $\mEl\B^\text{i}$, see Fig.~\ref{fig:multF}.
	However, it should be noted that configurations $\mEl\B^\text{i}$ of the material body which are such that $\mEl\te F^\text{i}=(\nablaX{}_\xi\ve\varphi^\text{i})^T$
	are the gradients of inelastic partial motion mappings $\ve {}_\xi\ve\varphi^\text{i}: \mathcal{B}_0 \times \mathcal T \to \mEl\mathcal{B}^i$ generally does not exist \cite[Sect.~1.10.3]{Haupt2000}, \cite[Sect.~14.3.1]{deSouzaNeto2008a}. Thus, the intermediate configuration concept is only valid in the local (pointwise) sense \cite[Sect.~14.3.1]{deSouzaNeto2008a}.
	Nevertheless, we define tensor quantities on the basis of $\mEl\te F^\text{e}$ and $\mEl\te F^\text{i}$ in analogy to the kinematic measures presented in Sect.~\ref{sec:kinematics_stress}. Note that these tensors can also be related to the fictitious intermediate configurations. Please also note that the decompositions of the deformation gradient are not unique, as the rotational parts remain undefined., i.e., $\te F = \mEl\te F^\text{e} \cdot \mEl\te F^\text{i}= \mEl\te F^\text{e} \cdot \mEl\te Q^T \cdot \mEl\te Q \cdot \mEl\te F^\text{i} = \mEl\te F^\text{e,*} \cdot \mEl\te F^\text{i,*}$ \cite{Haupt2000,Holthusen2024}. However, as we will not calculate quantities related to the intermediate configurations directly, this is by no means a problem.
\end{rmk}

Based on $\mEl\te F^\text{e}$ and $\mEl\te F^\text{i}$, we introduce the following related right Cauchy-Green deformation tensors and their isochoric, i.e., unimodular, parts:
\begin{align}
	\mEl\te C^\text{i} &= (\mEl\te F^\text{i})^T \cdot \mEl\te F^\text{i} \in \Sym \cap \GLp \text{ , } \mEl\bCi = (\mEl J^\text{i})^{-2/3}\mEl\te C^\text{i} \in\Sym \cap \SL \text{ and } \\
	\mEl\te C^\text{e} &= (\mEl\te F^\text{e})^T \cdot \mEl\te F^\text{e} \in \Sym \cap \GLp
	\text{ , } \mEl\bCe = (\mEl J^\text{e})^{-2/3}\mEl\te C^\text{e} \in \Sym \cap \SL \; .
\end{align} 
Here, the inelastic portions $\mEl\te C^\text{i}$ are related to the reference configuration, whereas the elastic portions $\mEl\te C^\text{e}$ are related to the fictitious inelastic intermediate configurations, cf. Remark~\ref{rmk:mult}. For the calculations applied later, we represent the elastic right Cauchy-Green deformation tensors and their isochoric parts
\begin{align}
	\mEl\te C^\text{e} = (\mEl\te F^\text{i})^{-T} \cdot \te C \cdot (\mEl\te F^\text{i})^{-1} \text{ and }
	\mEl \bCe = 
	(\mEl\bte F{}^\text{i})^{-T} \cdot \bte C \cdot (\mEl\bte F{}^\text{i})^{-1}
	= J^{-2/3} (\mEl J^\text{i})^{2/3}(\mEl\te F^\text{i})^{-T} \cdot \te C \cdot (\mEl\te F{}^\text{i})^{-1} \label{eq:Ce}
\end{align}
in terms of $\te C$ and $\mEl \te F^\text{i}$ by using Eq.~\eqref{eq:mult}.

Eq.~\eqref{eq:Ce} enables us to express the invariants $\mEl\bar I_1^\text{e}, \mEl\bar I_2^\text{e}\in\R_{\ge 0}$ in terms of $\bte C$ and the isochoric parts of the inelastic right Cauchy-Green deformation tensors $\mEl\bCi$ \cite{Rambausek2022,Ciambella2024}:
\begin{align}
	\mEl\bar I_1^\text{e} = \tr \mEl\bCe =  \bte C : (\mEl\bCi)^{-1} \text{ and }
	\mEl\bar I_2^\text{e} = \tr \left(\cof \mEl\bCe\right) =  \bte C{}^{-1} : \mEl\bCi \; . \label{eq:invariants_el}
\end{align}
Since we assume perfectly incompressible materials, i.e., $J = 1$, $\mEl I_3^\text{e} = \det \mEl\te C^\text{e}$ is not needed. However, it should be noted that the assumption of incompressibility does not imply $\mEl J^\text{e} = \mEl J^\text{i} = 1$. This has to be enforced additionally by the evolution equation if required, cf. Theorem~\ref{prop:unimodular}.

\subsubsection{Free energy and evaluation of the Clausius-Duhem inequality}

\begin{figure}
	\centering
	\includegraphics{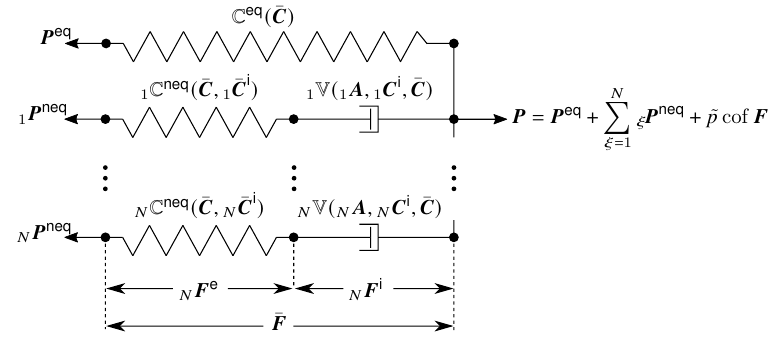}
	\caption{Rheological model of an incompressible generalized Maxwell model ($\det \te F = 1$) in finite strain viscoelasticity. The model consists of a spring for the equilibrium part and $N$ Maxwell elements. The tangents of equilibrium $\tttte C^\text{eq}(\bte C)$ and non-equilibrium $\mEl \tttte C^\text{neq}(\bte C, \mEl \bte C^\text{i})$ components may depend nonlinearly on the isochoric parts of deformation $\bte C$ and inelastic deformations $\mEl \bte C^\text{i}$, respectively. Similarly, the viscosity tensors  $\mEl \tttte V(\mEl \te A, \mEl \te C^\text{i}, \bte C)$ can depend nonlinearly on conjugate thermodynamic forces $\mEl \te A$ and $\mEl \te C^\text{i}$, $\bte C$. The pressure-like Lagrangian multiplier $\tilde p$ enforces incompressibility.}
	\label{fig:maxwell}
\end{figure}

\paragraph{Additive decomposition into equilibrium and non-equilibrium parts}
After discussing the multiplicative decomposition of the deformation gradient, we continue with the formulation of the free energy. 
As common in viscoelasticity, we assume an additive decomposition into equilibrium part $\psi^\text{eq}: \Sym\cap \SL \to \R_{\ge 0}, \bte C \mapsto \psi^\text{eq}(\bte C)$ and $N$ non-equilibrium parts $\mEl\psi^\text{neq}: \Sym\cap \SL \to \R_{\ge 0}, \mEl \bCe \mapsto \mEl \psi^\text{neq}(\mEl\bCe)$ depending on the elastic right Cauchy-Green deformation tensor \cite{Reese1998,Kumar2016,Rambausek2022,Ciambella2024}.
In addition, to enforce incompressibility, we add $\psi^\text{inc}: \R_{\ge 0} \times \R \to \R, (J,\tilde p) \mapsto \psi^\text{inc}(J,\tilde p) = \tilde p (J-1)$ depending on the Lagrange multiplier $\tilde p\in\R$ \cite[Sect.~6.3]{Holzapfel2000},\cite{Dammass2025a,Dammass2025b}. $\tilde p$ has to be determined from the boundary conditions when a boundary value problem is solved later. The rheological model of the incompressible model is depicted in Fig.~\ref{fig:maxwell}.

As we restrict ourselves to isotropy, we formulate the equilibrium and non-equilibrium potentials in terms of the invariants $\bar I_1 = \tr \bte C$, $\bar I_2 = \tr(\cof \bte C)$ and $\mEl\bar I_1^\text{e}(\bte C, \mEl\bCi)$, $\mEl\bar I_2^\text{e}(\bte C, \mEl\bCi)$ given in Eq.~\eqref{eq:invariants_el}, respectively. With that, the entire free energy density functional is defined as 
\begin{align}
	\psi(\te F, \mte C^\text{i},\tilde p) = \psi^\text{eq}(\bar I_1,\bar I_2) + \sum_{\xi=1}^{N}\mEl\psi^\text{neq}(\mEl \bar I_1^\text{e},\mEl \bar I_2^\text{e}) + \tilde p (J-1) \; ,
	\label{eq:psi}
\end{align}
where the tuple $\mte C^\text{i}:=({}_1\te C^\text{i}, {}_2\te C^\text{i}, \ldots, {}_N\te C^\text{i})$. 
Thus, $\te F$, $\mEl\te C^\text{i}$ and $\tilde p$ are chosen as \emph{independent constitutive variables}.
For brevity, the invariant sets are summarized in the tuples $\I^\text{eq}:=(\bar I_1,\bar I_2)\in\R^2_{\ge 0}$ as well as $\mEl\I^\text{neq}:=(\mEl\bar I_1^\text{e},\mEl\bar I_2^\text{e})\in\R^2_{\ge 0}$ and the invariant sets for all Maxwell elements in the tuple $\I^\text{neq}:=({}_1\!\I^\text{neq},{}_2\!\I^\text{neq}, \ldots, {}_N\!\I^\text{neq})$, respectively. With the choice of the invariants as arguments for the free energy, we ensure \emph{objectivity}, \emph{material symmetry}, and \emph{indifference to the choice of intermediate configuration}, i.e., \emph{invariance with respect to the rotational part} $\mEl \te R^\text{i} \in \SO$ of $\mEl \te F^\text{i} = \mEl \te R^\text{i} \cdot \mEl \te U^\text{i}$ \cite{Haupt2000,Sansour2007,Holthusen2025}.

\paragraph{Clausius-Duhem inequality}
In the following, we will discuss \emph{thermodynamic consistency} using the  \emph{Clausius-Duhem inequality (CDI)} $\mathcal D= \te P : \dot{\te F}-\dot \psi \ge 0$ with $\mathcal D$ being the dissipation rate. To do so, we assume the following functional dependencies:
\begin{align}
	\psi(\te F, \mte C^\text{i},\tilde p) \; , \; \te P(\te F, \mte C^\text{i},\tilde p) \text{ and } \mEl \dot{\te C}^\text{i} = \mEl \te f(\te F, \mte C^\text{i}) \; , \; \xi \in\{1,2,\ldots,N\} \; .
	\label{eq:ansatzes}
\end{align}
In the next step we apply the selected ansatzes~\eqref{eq:ansatzes} to the CDI and use the chain rule to obtain
\begin{align}
	\begin{aligned}
		\mathcal D = -\dot \psi + \te P : \dot{\te F} 
		&= \left(\te P-\diffp{\psi^\text{eq}}{\te F}-\sum_{\xi=1}^{N}\diffp{\mEl\psi^\text{neq}}{\te F}
		- \tilde p \cof \te F\right) : \dot{\te F}\\
		&- \sum_{\xi=1}^{N}2\diffp{\mEl\psi^\text{neq}}{\mEl\te C^\text{i}}: \frac{1}{2}\mEl \dot{\te C}{}^\text{i} \\
		&+ (J-1)\dot{\tilde p}
		\ge 0 \; \forall \dot{\te F}\in \Ln_2, \te F \in \GLp, \mEl \te C^\text{i}\in\Sym, \dot{\tilde p}, \tilde p\in\R \; .
		\label{eq:CDU}
	\end{aligned}
\end{align}
By applying the procedure of Coleman,~Noll~and~Gurtin \cite{Coleman1963,Coleman1967}, we find the three \emph{necessary and sufficient conditions} 
\begin{align}
	\te P = \diffp{\psi^\text{eq}}{\te F} + \sum_{\xi=1}^{N}\diffp{\mEl\psi^\text{neq}}{\te F}
	+ \tilde p \cof \te F \; \wedge \; J=1 
	\; \wedge \; 
	\mathcal D = 
	- \sum_{\xi=1}^{N}2\diffp{\mEl\psi^\text{neq}}{\mEl\te C^\text{i}}: \frac{1}{2}\mEl \dot{\te C}{}^\text{i}
	\ge 0 \; \forall \te F\in \SL, \mEl {\te C}{}^\text{i}\in\Sym 
	\label{eq:conditions}
\end{align}
from inequality~\eqref{eq:CDU}.
The first condition is the definition of the 1st Piola-Kirchhoff stress tensor that is a sum of the equilibrium stress $\te P^\text{eq}=\partial_{\te F}\psi^\text{eq}$, $N$ non-equilibrium stresses $\mEl \te P^\text{neq}=\partial_{\te F}\mEl\psi^\text{neq}$ and a pressure term $\te P^\text{inc}=\tilde p\cof \te F$.
The corresponding 2nd Piola-Kirchhoff and Cauchy stress tensors can be computed by applying push-forward operations, i.e., the inverses of the pull-backs given in Sect.~\ref{sec:kinematics_stress}. 	 
The second condition states the unimodularity of the deformation gradient, i.e., $\te F \in \SL$.
The third condition requires the dissipation rate $\mathcal D$ to be non-negative. To guarantee this, we have to define a specific form of the evolution equations for the inelastic deformations $\mEl \te C^\text{i}$ in the following.

\begin{rmk}
	\label{rmk:inc}
	Since the equilibrium and non-equilibrium energies as chosen in Eq.~\eqref{eq:psi} only depend on invariants of the isochoric tensors $\bte C$ and $\mEl\bCi$, we find that the Lagrange multiplier is minus the hydrostatic pressure, i.e., $\tilde p = - p =  1/3 \tr \te \sigma$, cf. \cite[App.~A.1]{Dammass2025a}. Furthermore, the isochoric invariants automatically guarantee that the stress tensors $\te P^\text{eq}$ and $\te P^\text{neq}$ are $\zero$ for $(\te F, \mEl \te C^\text{i})=(\one, \one)$.
\end{rmk}

\subsubsection{Dual dissipation potential and definition of evolution equations}

\paragraph{Thermodynamic consistency}
In order to fulfill the dissipation inequality, we make use of the approach called \emph{generalized standard materials (GSMs)} or \emph{two-potential framework} \cite{Kumar2016,Rambausek2022,Ciambella2024}. First, we define the stress-type \emph{thermodynamic forces} 
\begin{align}
	\mEl\te A := -2 \diffp{\mEl\psi^\text{neq}}{\mEl\te C^\text{i}} \in \Sym
\end{align}
that are dual to the deformation-type internal variables $\mEl\te C^\text{i}$.\footnote{In contrast to the linear theory, the thermodynamic forces $\mEl\te A$ are not equal to the non-equilibrium stresses $\mEl\te T^\text{neq}=2\partial_{\te C} \mEl\psi^\text{neq} = \te F^{-1}\cdot \partial_{\te F} \mEl\psi^\text{neq}$.}

Following the GSM approach, we introduce a so-called \emph{dual dissipation potential}
\begin{align}
	\phi^*: \Sym^N \times (\Sym\cap\GLp)^N \times \Sym \cap \SL \to \R_{\ge 0}, (\mte A,\mte C^\text{i},\bte C) \mapsto \phi^*(\mte A,\mte C^\text{i},\bte C) = \sum_{\xi=1}^{N}\mEl\phi^*(\mEl\te A, \mEl\te C^\text{i},\bte C) 
	\label{eq:dissi_pot}
\end{align}
with the tuple $\mte A:=({}_1\te A, {}_2\te A, \ldots, {}_N\te A)$
and define the evolution equations 
\begin{align}
	\mEl\dot{\te C}{}^\text{i} = 2\diffp{\mEl\phi^*}{\mEl\te A}
	\;, \; \xi \in\{1,2,\ldots,N\}  \label{eq:BiotEquation}
\end{align}
that are nonlinear systems of \emph{ordinary differential equations (ODEs)} in time.
Thus, the dissipation rate follows to
\begin{align}
	\mathcal D = \sum_{\xi=1}^{N} \mEl \te A : \diffp{\mEl\phi^*}{\mEl\te A}
\end{align}
which is always guaranteed to be non-negative if the potentials $\mEl\phi^*(\mEl\te A, \mEl\te C^\text{i},\bte C)$ are convex in $\mEl\te A$ on the convex set $\Sym$ and it holds $\mEl\phi^*(\zero, \mEl\te C^\text{i},\bte C)=0 \; \wedge \; \mEl\phi^*(\mEl\te A, \mEl\te C^\text{i},\bte C)\ge 0 \; \forall \mEl\te A, \mEl\te C^\text{i}, \bte C\in \Sym$.\footnote{\label{foot:conditions_phi}
	Note that the conditions
	\begin{align*}
		\mEl\phi^*(\mEl\te A, \mEl\te C^\text{i},\bte C) \text{ convex w.r.t. } \mEl \te A \; \wedge \;
		\mEl\phi^*(\zero, \mEl\te C^\text{i},\bte C)=0 \; \wedge \;
		\mEl\phi^*(\mEl\te A, \mEl\te C^\text{i},\bte C)\ge 0 \; \forall \mEl\te A, \mEl\te C^\text{i}, \bte C\in \Sym
	\end{align*}
	are equivalent to
	\begin{align*}
		\mEl\phi^*(\mEl\te A, \mEl\te C^\text{i},\bte C) \text{ convex w.r.t. } \mEl \te A \; \wedge \;
		\mEl\phi^*(\zero, \mEl\te C^\text{i},\bte C)=0 \; \wedge \;
		\partial_{\mEl\te A}\mEl\phi^*|_{(\zero,\mEl\te C^\text{i}, \bte C)} = \zero \; \forall \mEl\te C^\text{i}, \bte C\in \Sym \; .
\end{align*}}
The evolution equations~\eqref{eq:BiotEquation} following from the GSM approach are a special choice for $\mEl \dot{\te C}^\text{i} = \mEl \te f(\te F, \mte C^\text{i},\tilde p)$. Thus, Eq.~\eqref{eq:BiotEquation} in combination with stated requirements for the dual dissipation potential are therefore only \emph{sufficient} for $\mathcal D\ge 0$ and not necessary and sufficient.

\begin{rmk}
	\label{rmk:biot_standard}
	Another option to ensure $\mathcal D\ge 0$ is to introduce the dissipation potential 
	$\phi: \Sym^N \times (\Sym\cap\GLp)^N \times \Sym\cap\SL \to \R_{\ge 0}, (\dot{\mte C}{}^\text{i},\mte C^\text{i},\bte C) \mapsto \phi(\dot{\mte C}{}^\text{i},\mte C^\text{i},\bte C)$ equivalent to Eq.~\eqref{eq:dissi_pot} and to set $\mEl \te A = 2\partial_{\mEl \dot{\te C}{}^\text{i}} \mEl \phi$, which gives the evolution equations
	\begin{align}
		\diffp{\mEl\psi^\text{neq}}{\mEl\te C^\text{i}} + \diffp{\mEl \phi}{\mEl \dot{\te C}{}^\text{i}} = \zero \; ,
		\label{eq:biot_orig}
	\end{align}
	cf. \cite{Rambausek2022,Ciambella2024}. If $\phi(\dot{\mte C}{}^\text{i},\mte C^\text{i},\bte C)$ is convex w.r.t. $\dot{\mte C}{}^\text{i}$ and $\phi(\zero,\mte C^\text{i},\bte C)=0 \wedge \phi(\dot{\mte C}{}^\text{i},\mte C^\text{i},\bte C) \ge 0 \, \forall \dot{\mte C}{}^\text{i},\mte C^\text{i},\bte C \in \Sym$ it holds $\mathcal D\ge 0$.
	The dual dissipation potential and the dissipation potential are linked via the \emph{Legendre-Fenchel transformation} 
	\begin{align}
		\phi^*(\mte A,\mte C^\text{i},\bte C) = \underset{\dot{\mte C}{}^\text{i}\in\Sym^N}{\sup}\left(
		\frac{1}{2}\mte A \bullet \dot{\mte C}{}^\text{i} -\phi(\dot{\mte C}{}^\text{i},\mte C^\text{i},\bte C)
		\right) \; ,
		\label{eq:legendre-fenchel}
	\end{align}
	with $\bullet$ denoting the double contraction of the tuple elements \cite{Miehe2011a}. 
\end{rmk}

\paragraph{Unimodular inelastic deformation}
In order to guarantee that $\mEl J^\text{i} = 1$ applies during the evolution of the internal variables, i.e., the inelastic deformations $\mEl \te C^\text{i}$ are unimodular, the following specific structure of the dual dissipation potential is chosen: 
\begin{align}
	\mEl \phi^*: \Sym \times \Sym\cap\GLp \times \Sym\cap\SL \to \R_{\ge 0}, (\mEl \te A, \mEl \te C^\text{i}, \bte C) \mapsto \mEl \phi^*(\mEl \te A^\text{p}(\mEl \te A,\mEl \te C^\text{i}), \bte C) \; .
	\label{eq:mod_diss}
\end{align}
Thereby, $\mEl \te A^\text{p}$ are the \emph{projected thermodynamic forces} defined via 
\begin{align}
	\mEl \te A^\text{p} := \mEl\tttte P:\mEl \te A  =  \mEl \te A - \frac{1}{3} \left(\mEl\te C^\text{i}: \mEl\te A\right) (\mEl \te C^\text{i})^{-1} \in \Sym
	\;, \; \xi \in\{1,2,\ldots,N\} 
	\label{eq:projection}
\end{align}
with the projectors  $\mEl\tttte P := \ttttes 1^\text{s} - 1/3 (\mEl \te C^\text{i})^{-1} \otimes \mEl \te C^\text{i} \in \Ln_4$ of 4th order.

\begin{lemma}
	\label{lemma:projector}
	Let $\phi^*(\mEl \te A^\text{p}(\mEl \te A,\mEl \te C^\text{i}), \bte C)$ the dual dissipation potential according to Eq.~\eqref{eq:mod_diss} and let $\xi \in\{1,2,\ldots,N\}$.
	Then it holds $\partial_{\mEl \te A} \mEl \phi^* : (\mEl \te C^\text{i})^{-1} = 0$.
\end{lemma}

\begin{proof}
	By using the chain rule it follows
	\begin{align}
		\diffp{\mEl\phi^*}{\mEl\te A^\text{p}} :
		\diffp{\mEl\te A^\text{p}}{\mEl\te A} : (\mEl \te C^\text{i})^{-1}
		&=
		\diffp{\mEl\phi^*}{\mEl\te A^\text{p}}:\left[\ttttes 1^\text{s} - \frac{1}{3}(\mEl \te C^\text{i})^{-1} \otimes \mEl \te C^\text{i}\right] : (\mEl \te C^\text{i})^{-1}
		\nonumber \\
		&= 
		\diffp{\mEl\phi^*}{\mEl\te A^\text{p}}:\left[(\mEl \te C^\text{i})^{-1} - (\mEl \te C^\text{i})^{-1}\right] = 0
		\; .
	\end{align}
\end{proof}

\begin{rmk}
	\label{rmk:projection}
	Note that the projections~\eqref{eq:projection} still guarantee that the potentials $\mEl\phi^*(\mEl\te A^\text{p}(\mEl \te A, \mEl \te C^\text{i}),\bte C)$ are convex in $\mEl \te A$ as long as $\mEl\phi^*(\mEl\te A^\text{p}(\mEl \te A, \mEl \te C^\text{i}),\bte C)$ is convex w.r.t. $\mEl \te A^\text{p}$ for all $\xi\in\{1,2,\ldots,N\}$. This applies because $\mEl\tttte P:\mEl \te A$ are linear mappings, cf. \ref{app:invariants_dissipation}, Proposition~\ref{prop:projection_convex}.
\end{rmk}

\begin{theorem}
	\label{prop:unimodular}
	Let $\mEl \phi^*(\mEl \te A^\text{p}(\mEl \te A,\mEl \te C^\text{i}), \bte C)$, $\xi \in\{1,2,\ldots,N\}$ the dual dissipation potential according to Eq.~\eqref{eq:mod_diss} and let $\mEl \te C^\text{i}=\one$ at $t=t_0$.
	Furthermore let the evolution be defined by $\mEl\dot{\te C}{}^\text{i} = 2\partial_{\mEl \te A} \mEl \phi^*$.
	Then the inelastic deformations $\mEl \te C^\text{i}$ stay always unimodular, i.e., $\det \mEl \te C^\text{i} = 1$.
\end{theorem}

\begin{proof}
	As $\left(\frac{\dx}{\dx t}\det \mEl \te C^\text{i} = 0 \; \forall t\ge t_0 \, \wedge\,
	\mEl \te C^\text{i} = \one \text{ at } t_0
	\right)$ implies $\det \mEl \te C^\text{i} = 1 \; \forall t\ge t_0$, it is sufficient to prove that $\frac{\dx}{\dx t} \det \mEl \te C^\text{i} = 0 \; \forall t\ge t_0$ holds.
	These conditions can be rewritten as
	\begin{align}
		\difft{}{t} \det \mEl\te C^\text{i} = \det \left(\mEl\te C^\text{i}\right) (\mEl \te C^\text{i})^{-1} : \mEl\dot{\te C}{}^\text{i} = 0 \; ,
		\label{eq:inc}
	\end{align}
	cf. \cite{Ciambella2025}. By inserting the evolution equations~\eqref{eq:BiotEquation}, we find 
	\begin{align}
		\mEl\dot{\te C}{}^\text{i}:(\mEl \te C^\text{i})^{-1}=2\partial_{\mEl \te A}\mEl\phi^* : (\mEl \te C^\text{i})^{-1} = 0 \; ,
	\end{align}
	which holds true by using Lemma~\ref{lemma:projector}. 
\end{proof}

\paragraph{Isotropy}
Since we restrict ourselves to isotropy, the dissipation potential, similar to the free energy density, has to be an \emph{isotropic tensor function}, i.e., $\mEl\phi^*(\mEl\te A^\text{p},\bte C) = \mEl\phi^*(\te Q \cdot \mEl\te A^\text{p}\cdot \te Q^T,\te Q \cdot \bte C\cdot \te Q^T) \; \forall \te Q\in \Othree$.

To this end, we can build an \emph{irreducible functional basis}, i.e., a complete and irreducible invariant set, by using the procedure according to Boehler~\cite{Boehler1977}. However, since convexity w.r.t. $\mEl\te A^\text{p}$ is required, cf. Remark~\ref{rmk:projection}, we follow Rosenkranz~el~al.~\cite{Rosenkranz2024} and replace the cubic invariant in $\mEl\te A^\text{p}$ with a quartic one.\footnote{It is worth noting that the invariant set $\mEl \I^{\phi^*}$ given in Eq.~\eqref{eq:invariants_phi} does not form an irreducible integrity basis, as $\tr (\mEl \te A^\text{p})^3$ cannot be represented as a polynomial in $\mEl \I^{\phi^*}$. This can be shown by using the Cayley-Hamilton theorem:
	\begin{align*}
		\tr (\mEl \te A^\text{p})^3 = \frac{4}{3}\left(
		4 \frac{\mEl I_3^{\phi^*}}{\mEl I_1^{\phi^*}}  + 2 I_1^{\phi^*} I_2^{\phi^*} -  2 \frac{(I_2^{\phi^*})^2}{ I_1^{\phi^*}}-\frac{1}{6} (I_1^{\phi^*})^3 
		\right) \; .
	\end{align*}
	However, $\mEl \I^{\phi^*}$ forms a functional basis.
}
With that, we find the invariant sets consisting of 
\begin{align}
	\begin{split}
		\mEl I_1^{\phi^*} &= \tr \mEl \te A^\text{p}, \; 
		\mEl I_2^{\phi^*} = \frac{1}{2}\tr  \left(\mEl\te A^\text{p}\right)^2, \;
		\mEl I_3^{\phi^*} = \frac{1}{4}\tr  \left(\mEl\te A^\text{p}\right)^4,\;
		\mEl I_4^{\phi^*} = \tr  \bte C,\;
		\mEl I_5^{\phi^*} = \frac{1}{2}\tr  \bte C^2,\;\\
		\mEl I_6^{\phi^*} &= \tr  \left(\mEl \te A^\text{p}\cdot \bte C\right),\;
		\mEl I_7^{\phi^*} = \frac{1}{2}\tr  \left((\mEl \te A^\text{p})^2\cdot \bte C\right),\;
		\mEl I_8^{\phi^*} = \tr  \left(\mEl \te A^\text{p}\cdot \bte C^2\right),\;
		\mEl I_9^{\phi^*} = \frac{1}{2}\tr  \left((\mEl \te A^\text{p})^2\cdot \bte C^2\right) \; ,
		\label{eq:invariants_phi}
	\end{split}
\end{align}
$\xi\in\{1,2,\ldots,N\}$, where we collect the set for each $\xi$ in the tuple $\mEl \I^{\phi^*}$.
The convexity of the mixed invariants is proven in \ref{app:invariants_dissipation}. For brevity, we also introduce the tuple $\I^{\phi^*}:=({}_1\!\I^{\phi^*},{}_2\!\I^{\phi^*}, \ldots, {}_N\!\I^{\phi^*})$.

\subsection{Linearization of the model for small strains}
\label{sec:linearization}

In this subsection, we discuss the reduction of the presented finite viscoelasticity theory to linear viscoelasticity at small strains and prove consistency with these well-known model equations.
To this end, we carry out \emph{Taylor series} expansions of the potentials up to the second order.

\subsubsection{Equilibrium energy of the free energy}
We start with the equilibrium energy depending on the two isochoric invariants $\bar I_1,\bar I_2$ of $\bte C$. The Taylor series gives
\begin{align}
	\mathcal T_{\one}\psi^\text{eq} =
	\psi^\text{eq}(\I^\text{eq})\big|_{\one} + \diffp{\psi^\text{eq}}{\te C}\Big|_{\one} : (\te C - \one)
	+ \frac{1}{2}(\te C - \one):\diffp{{}^2\psi^\text{eq}}{\te C\partial \te C}\Big|_{\one} : (\te C - \one) + \mathrm{HOT}\; ,
\end{align}
where $\mathcal T_{\te 1}$ denotes the Taylor series expansion at $\te C=\one$ and $\mathrm{HOT}$ are higher order terms.
Accounting for the structure of the isochoric invariants, the equation above reduces to
\begin{align}
	= \frac{1}{2}\te E:\underbrace{4\diffp{{}^2\psi^\text{eq}}{\te C\partial \te C}\Big|_{\one}}_{=:\tttte C^\text{eq}} : \te E + \mathrm{HOT} 
	\quad \text{with} \quad
	\tttte C^\text{eq} = \underbrace{4\sum_{\alpha=1}^{2} \diffp{\psi^\text{eq}}{\bar I_\alpha}\Bigg|_{\one}}_{=:2\mu}\underbrace{\left(\ttttes 1^\text{s} - \frac{1}{3} \te 1 \otimes \te 1\right)}_{=:\tttte P^\text{d}} 
	\label{eq:lin_equilibrium}
\end{align}
as $\partial_{\te C} \bar I_1|_{\one} = \partial_{\te C} \bar I_2|_\one = \zero$ and the equilibrium energy is assumed to vanish for $\te C=\one$.\footnote{The Hessian of the equilibrium energy is given by
	\begin{align*}
		\diffp{{}^2\psi^\text{eq}}{\te C\partial \te C} = \sum_{\alpha=1}^{2}\sum_{\beta=1}^{2}
		\diffp{{}^2\psi^\text{eq}}{\bar I_\alpha\partial \bar I_\beta} \diffp{\bar I_\alpha}{\te C}
		\otimes 
		\diffp{\bar I_\beta}{\te C}
		+\sum_{\alpha=1}^{2} \diffp{\psi^\text{eq}}{\bar I_\alpha} \diffp{{}^2 \bar I_\alpha}{\te C\partial \te C} \in \Sym_4 \; .
	\end{align*}
}
In the equation above, $\tttte P^\text{d} \in \Sym_4$ is the 4th order deviator projector and $\mu\in\R_{>0}$ the initial shear modulus. Thus, after a geometric linearization of $\te E$, that gives the technical strain $\te \varepsilon = 1/2(\nabla \ve u + (\nabla \ve u)^T) \in \Sym$, we get the well-known equilibrium energy
$\psi^\text{eq} = 1/2 \te \varepsilon : \tttte C^\text{eq} : \te \varepsilon$ with the constant tangent modulus $\tttte C^\text{eq} = 2\mu \tttte P^\text{d}\in\Sym_4$.

\subsubsection{Non-equilibrium energy of the free energy}
For the non-equilibrium energy, we form a \emph{Taylor series} up to quadratic order in $\te C$ and the inelastic deformation tensors $\mEl \te C^\text{i}$:
\begin{align}
	\begin{split}
		\mathcal T_{(\one,\one,\ldots,\one)}\psi^\text{neq} &=
		\sum_{\xi=1}^{N}\mEl\psi^\text{neq}(\mEl\I^\text{neq})\big|_{(\one,\one)} + 
		\sum_{\xi=1}^{N}\diffp{\mEl\psi^\text{neq}}{\te C}\Big|_{(\one,\one)} : (\te C - \one) +
		\sum_{\xi=1}^{N}\diffp{\mEl\psi^\text{neq}}{\mEl\te C^\text{i}}\Big|_{(\one,\one)} : (\mEl\te C^\text{i} - \one)  \\
		&+ \sum_{\xi=1}^{N}\frac{1}{2}(\te C - \one):\diffp{{}^2\mEl\psi^\text{neq}}{\te C\partial \te C}\Big|_{(\one,\one)} : (\te C - \one)
		+ \sum_{\xi=1}^{N}(\te C - \one):\diffp{{}^2\mEl\psi^\text{neq}}{\te C\partial \mEl\te C^\text{i}}\Big|_{(\one,\one)} : (\mEl\te C^\text{i} - \one) \\
		&+ \sum_{\xi=1}^{N}\frac{1}{2}(\mEl\te C^\text{i} - \one):\diffp{{}^2\mEl\psi^\text{neq}}{\mEl\te C^\text{i}\partial \mEl\te C^\text{i}}\Big|_{(\one,\one)} : (\mEl\te C^\text{i} - \one) + \mathrm{HOT} \; . \label{eq:lin_neq_1}
	\end{split}
\end{align}
As $\partial_{\te C} \mEl \bar I_1^\text{e}|_{(\one,\one)} = \partial_{\te C} \mEl \bar I_2^\text{e}|_{(\one,\one)} = \zero$, $\partial_{\mEl \te C^\text{i}} \mEl \bar I_1^\text{e}|_{(\one,\one)} =\partial_{\mEl \te C^\text{i}} \mEl \bar I_2^\text{e}|_{(\one,\one)} = \zero$ and the non-equilibrium energy is assumed to vanish for $\te C=\one$ and $\mEl\te C^\text{i}=\one$, the evaluation of Eq.~\eqref{eq:lin_neq_1} yields
\begin{align}
	\mathcal T_{(\one,\one,\ldots,\one)}\psi^\text{neq} &= \frac{1}{2}\sum_{\xi=1}^{N}\left[
	\te E : \mEl\tttte C^\text{neq} : \te E -
	2 \te E : \mEl\tttte C^\text{neq} : \mEl\te E^\text{i} +
	\mEl\te E^\text{i} : \mEl\tttte C^\text{neq} : \mEl\te E^\text{i}
	\right] + \mathrm{HOT} \nonumber \\
	&= \frac{1}{2}\sum_{\xi=1}^{N}
	(\te E - \mEl\te E^\text{i}) : \mEl\tttte C^\text{neq} : (\te E - \mEl\te E^\text{i}) + \mathrm{HOT}
	\; , 
\end{align}
with
\begin{align}
	\mEl\tttte C^\text{neq} 
	:= 4\diffp{{}^2\mEl\psi^\text{neq}}{\te C\partial \te C}\Bigg|_{(\one,\one)} 
	= -4\diffp{{}^2\mEl\psi^\text{neq}}{\te C\partial \mEl\te C^\text{i}}\Bigg|_{(\one,\one)}
	= 4\diffp{{}^2\mEl\psi^\text{neq}}{\mEl\te C^\text{i}\partial \mEl\te C^\text{i}}\Bigg|_{(\one,\one)}
	= \underbrace{4\sum_{\alpha=1}^{2} \diffp{\psi^\text{neq}}{\mEl \bar I_\alpha^\text{e}}\Bigg|_{(\one,\one)}}_{=:2\mEl\mu}\tttte P^\text{d} \in \Sym_4 \; .
	\label{eq:lin_nonequilibrium}
\end{align}
Therein, $\mEl \te E^\text{i} := 1/2(\mEl \te C^\text{i} - \one) \in \Sym$ is the inelastic Green-Lagrange strain tensor. Geometric linearization of $\te E$ and $\mEl \te E^\text{i}$ gives $\te \varepsilon$ and $\mEl \te \varepsilon^\text{i}$, respectively. Thus, we get the non-equilibrium energy expressions 
$\mEl\psi^\text{neq}(\te \varepsilon,\mEl\te\varepsilon^\text{i}) = 1/2 (\te \varepsilon- \mEl \te\varepsilon^\text{i}) : \mEl\tttte C^\text{neq} : (\te \varepsilon-\mEl \te \varepsilon^\text{i})$ with the additive split of the strain $\te \varepsilon = \mEl \te \varepsilon^\text{e} + \mEl \te \varepsilon^\text{i}$ into elastic and inelastic parts and the constant tangent modules $\mEl\tttte C^\text{neq} = 2\mEl\mu_0 \tttte P^\text{d}$.\footnote{
	Note that the transition of the multiplicative decomposition of the deformation gradient to the additive decomposition of $\te \varepsilon$ also follows from a linearization of Eq~\eqref{eq:mult}, cf. \cite{Casey1985}.}
$\mEl \mu \in \R_{> 0}$ is the initial shear modulus corresponding to the $\xi$th Maxwell element.

\subsubsection{Incompressibility part  of the free energy}
Finally, the energy contribution to enforce the model's incompressibility is considered. The Taylor series gives
\begin{align}
	\mathcal T_{\one}\psi^\text{inc} &=
	\psi^\text{inc}(J,\tilde p)\big|_{(\one,\tilde p)} 
	+ \diffp{\psi^\text{inc}}{\te C}\Big|_{(\one,\tilde p)} : (\te C - \one)
	+ \frac{1}{2}(\te C - \one):\diffp{{}^2\psi^\text{inc}}{\te C\partial \te C}\Big|_{(\one,\tilde p)} : (\te C - \one) + \mathrm{HOT} \nonumber \\
	&= \tilde p\left(
	\tr \te E + \frac{1}{2} \tr^2 \te E - \te E : \te E
	\right) + \mathrm{HOT} \; .
\end{align}
Through a geometric linearization step of $\te E$, which again yields $\te \varepsilon$, and taking into account $\| \te \varepsilon\| \ll 1$, the quadratic terms disappear in relation to the linear term and we obtain $\psi^\text{inc}(\te \varepsilon, \tilde p) = \tilde p \tr \te \varepsilon$. 

Due to the deviatoric nature of the linearized equilibrium and non-equilibrium potentials, only $\psi^\text{inc}(\te \varepsilon, \tilde p)$ gives non-deviatoric contributions to the stress $\te \sigma = \partial_{\te \varepsilon} \psi$ and it follows $\tilde p = -p = 1/3 \tr \te \sigma$. 

\subsubsection{Dual dissipation potential}

After discussing the three contributions of the free energy, we consider now the dual dissipation potential consisting of $\mEl \phi^*(\mEl\I^{\phi^*})$ with $\mEl\I^{\phi^*}=\mEl\I^{\phi^*}(\mEl \te A^\text{p}, \bte C)$ and $\mEl \te A^\text{p} = \mEl \te A^\text{p}(\mEl \te A, \mEl \te C^\text{i})$.
The Taylor series expansion of the potentials $\mEl \phi^*(\mEl\I^{\phi^*})$ w.r.t. $\mEl \te A$ at $(\mEl \te A, \mEl \te C^\text{i}, \bte C) = (\zero, \one, \one)$ gives\footnote{The Taylor series expansion is only performed for $\mEl \te A$, and thus constant deformation and inelastic deformation, since it is assumed that the dual dissipation potential in the linear setting depends only on the thermodynamic forces.}
\begin{align}
	\mathcal T_{(\zero,\ldots,\zero,\one,\ldots,\one,\one)}\phi^* &=
	\sum_{\xi=1}^{N}\mEl\phi^*(\mEl\I^{\phi^*})\big|_{(\zero,\one,\one)} + 
	\sum_{\xi=1}^{N}\diffp{\mEl\phi^*}{\mEl\te A}\Bigg|_{(\zero,\one,\one)} : \mEl \te A 
	+ \sum_{\xi=1}^{N}\frac{1}{2}\mEl\te A:\diffp{{}^2\mEl\phi^*}{\mEl\te A\partial \mEl\te A}\Bigg|_{(\zero,\one,\one)} : \mEl\te A
	+ \mathrm{HOT} \\
	&= \frac{1}{2} \sum_{\xi=1}^{N}  \mEl \te A : \mEl\tttte V^{-1} : \mEl \te A + \mathrm{HOT}\; ,
\end{align}
with
\begin{align}
	\mEl\tttte V^{-1} :=\diffp{{}^2\mEl\phi^*}{\mEl\te A\partial \mEl\te A}\Bigg|_{(\zero,\one,\one)}
	= \underbrace{\sum_{\alpha\in\{2,7,9\}} \diffp{\mEl\phi^*}{\mEl I_\alpha^{\phi^*}}
		\Bigg|_{(\zero,\one,\one)}}_{=:1/(2\mEl\eta)} \tttte P^\text{d} \in \Sym_4 \; . 
	\label{eq:lin_dissipation}
\end{align}
Therein, it was used that $\partial_{\mEl \te A} \mEl I_\alpha^{\phi^*}|_{(\zero,\one,\one)} = \zero$ for all $\alpha\in\{1,2,\ldots,9\}$. After replacing $\mEl \te A$ with $\mEl \te a\in\Sym$ for clarity, we get $\mEl\phi^*(\mEl\te a) = 1/2 \mEl\te a : \mEl\tttte V^{-1} : \mEl \te a$ with the inverse viscosity tensor $\mEl\tttte V^{-1} = \frac{1}{2\mEl\eta} \tttte P^\text{d}\in\Sym_4$, where $\mEl \eta>0$ is the initial viscosity of the $\xi$th Maxwell element. 

\begin{rmk}
	\label{remark:pos_initial_params}
	When formulating the potentials of the finite viscoelasticity model, care should be taken to ensure that the derivatives w.r.t. the invariant sets $\I^\text{eq}$, $\I^\text{neq}$ and $\I^{\phi^*}$ are positive in the undeformed state, i.e., $(\te F, \mEl \te C^\text{i}, \mEl \te A) = (\one,\one,\zero)$, respectively. This is important to  guarantee a non-negative initial shear modulus $\mu$ of the equilibrium part as well as non-negative initial shear modules $\mEl \mu$ and viscosities $\mEl \eta$ of the Maxwell elements.
\end{rmk}

\subsection{Time discretization}
\label{sec:time_disc}

To solve the nonlinear ODEs~\eqref{eq:BiotEquation}, finite differences are used for time discretization. Thereby, we will make use of \emph{exponential integrators} \cite{deSouzaNeto2008a,Kumar2016,Holthusen2024} in order to construct an algorithm that preserves the unimodularity of the inelastic deformations, i.e., $\mEl \te C^\text{i} \in \Sym \cap \SL$, and is thus consistent to our model, which inherently guarantees this property, cf. Theorem~\ref{prop:unimodular}.

\subsubsection{Exponential map integrator}
By using the dual dissipation potential as defined in Eq.~\eqref{eq:mod_diss}, i.e., $\mEl\phi^*(\mEl\te A^\text{p}(\mEl\te A,\mEl\te C^\text{i}),\bte C)$ with $\mEl \te A^\text{p}$ according to Eq.~\eqref{eq:projection}, and evaluating the evolution equations $\mEl\dot{\te C}{}^\text{i} = 2\partial_{\mEl\te A}{\mEl\phi^*}$, $\xi \in\{1,2,\ldots,N\}$, we get
\begin{align}
	\mEl\dot{\te C}{}^\text{i} = 
	2\diffp{\mEl\phi^*}{\mEl\te A} = 
	2\diffp{\mEl\phi^*}{\mEl\te A^\text{p}} : \mEl\tttte P 
	= 
	\underbrace{
		2 \left(
		\diffp{\mEl\phi^*}{\mEl\te A^\text{p}} \cdot (\mEl \te C^\text{i})^{-1} - \frac{1}{3} \left( \diffp{\mEl\phi^*}{\mEl\te A^\text{p}} : (\mEl \te C^\text{i})^{-1}\right) \one
		\right)}_{=:\mEl \te H} \cdot \mEl \te C^\text{i} \; ,
	\label{eq:defH}
\end{align} 
where $\mEl\tttte P = \mEl\tttte P \cdot (\mEl \te C^\text{i})^{-1} \cdot \mEl \te C^\text{i}$ has been used. As can be seen from Eq.~\eqref{eq:defH}, the 2nd order tensors $\mEl\te H\in\Devi$ are deviatoric, i.e., $\tr \mEl \te H=0$ and in general non-symmetric, i.e., $\mEl \te H \ne \mEl \te H^T$. The form $\mEl\dot{\te C}{}^\text{i} = \mEl \te H \cdot \mEl \te C^\text{i}$ enables us to use an \emph{exponential integrator} \cite{deSouzaNeto2008a} for the numerical solution within the time interval $t\in[{}^{n-1}t,{}^{n}t]$ with $n\in\mathbb N$ and $n-1$ being the indices of the current and previous time step, respectively, and ${}^n \!\Delta t := {}^n t - {}^{n-1}t$ the $n$th time step width. By marking the time step that a tensor belongs to with an index in the upper left, we obtain 
\begin{align}
	\mElt{n}\te C{}^\text{i} = \exp\left(\mElt{n} \te H \, {}^n \!\Delta t\right) \cdot \mEltl{n-1} \te C^\text{i} \; ,
	\label{eq:exp_orig}
\end{align}
which automatically yields $\det \mElt{n}\te C{}^\text{i} = 1$, since $\tr \mElt{n}\te H = 0$ \cite[App.~B.1.1]{deSouzaNeto2008a} and $(\mElt{n}\te C{}^\text{i})^T=\mElt{n}\te C{}^\text{i}$ for the solution, cf. \ref{app:exp_integrators}, Theorem~\ref{theorem:sym_exp_map_orig}. However, due to the non-symmetric structure of $\mElt{n} \te H$, it is not guaranteed that $\mElt{n}\te C{}^\text{i}\in\Sym$ holds during  iterative solution, e.g., via a Newton-Raphson scheme.
Thus, we postulate the modified exponential integrator
\begin{align}
	\mElt{n}\te C{}^\text{i} = \sqrt{\mEltl{n-1} \te C^\text{i}} \cdot \exp\left(\mElt{n} \hat{\te H} \, {}^n \!\Delta t\right) \cdot \sqrt{\mEltl{n-1} \te C^\text{i}} \text{ with }
	\mElt{n}\hat{\te H}:=\sym\left(\sqrt{(\mEltl{n-1} \te C^\text{i})^{-1}} \cdot \mElt{n} \te H \cdot \sqrt{\mEltl{n-1} \te C^\text{i}}\right) \; .
	\label{eq:exp_mod}
\end{align}
which is, similar to the integrator~\eqref{eq:exp_orig}, an exact solution of the ODE~\eqref{eq:defH} for $\mElt{n} {\te H} = \text{const.}$, see \ref{app:exp_integrators}, Theorem~\ref{theorem:exact_exp_map_mod} for a proof. This modified exponential mapping also guarantees symmetry of $\mElt{n} \te C^\text{i}$ during the iterative solution.

\begin{rmk}
	Within our implementation, the tensor exponential, defined in Eq.~\eqref{eq:tensor_exponential}, is computed using TensorFlow's \verb|tf.linalg.expm|, which uses a combination of the scaling and squaring method and the Pad\'{e} approximation, cf. \cite{Higham2005} for details. Similarly, the tensor square root is computed via TensorFlow's \verb*|tf.linalg.sqrtm|, which uses the algorithm described in \cite{Higham1987}.
\end{rmk}

\subsubsection{Solution via Newton-Raphson scheme}

\begin{algorithm}
	\begin{small}
		\label{alg:Newton-Raphson}
		\caption{Solution of the time discretized evolution equations~\eqref{eq:exp_mod}.}
		\textbf{Initial guess: }$\mEltll{n,1}\te C{}^\text{i} = \exp\left(\mEltl{n-1} \te H \, {}^n \!\Delta t\right) \cdot \mEltl{n-1} \te C^\text{i}$ \tcp*{Explicit scheme for initialization}
		
		$j = 1$ 
		
		\While{$j\le n_\mathrm{iter} \wedge \|\mEltll{n,j}\te R\|>\mathrm{tol}$}{
			$\mEltll{n,j}\te A = 2\partial_{\mEltll{n,j}\te C^\text{i}} \mEl \psi^\text{neq}$ \tcp*{Compute thermodynamic forces}
			
			$\mEltll{n,j}\te H = 2\partial_{\mEltll{n,j}\te A} \mEl \phi^{*} \cdot (\mEltll{n,j}\te C{}^\text{i})^{-1}$  
			
			$\mEltll{n,j}\hat{\te H}=\sym\left(\sqrt{(\mEltl{n-1} \te C^\text{i})^{-1}} \cdot \mEltll{n,j} \te H \cdot \sqrt{\mEltl{n-1} \te C^\text{i}}\right)$
			
			$\mEltll{n,j}\te R = \mEltll{n,j}\te C{}^\text{i} - \sqrt{\mEltl{n-1} \te C^\text{i}} \cdot \exp\left(\mElt{n} \hat{\te H} \, {}^n \!\Delta t\right) \cdot \sqrt{\mEltl{n-1} \te C^\text{i}}$  \tcp*{Compute residuum}
			
			$\mEltll{n,j}\tttte K = \partial_{\mEltll{n,j}\te C^\text{i}}{\mEltll{n,j}\te R}$ \tcp*{Compute tangent}
			
			$\mEltll{n,j}\vect R \gets \mEltll{n,j}\te R\; $; 
			$\; \mEltll{n,j}\matr K \gets \mEltll{n,j}\tttte K$ \tcp*{Transform to Kelvin-Mandel}
			
			\textbf{Solve } $\mEltll{n,j}\matr K \, \mEltll{n,j}\Delta\vect C^\text{i}= -\mEltll{n,j}\vect R$ for $\xi\in\{1,2,\ldots, N\}$ \tcp*{Solve systems of equations}
			
			$\mEltll{n,j}\Delta\te C^\text{i}\gets\mEltll{n,j}\Delta\vect C^\text{i}$ \tcp*{Transform back to tensor notation}
			
			\textbf{Update } $\mEltl{n,j+1}\te C^\text{i} = \mEltll{n,j}\te C^\text{i} + \mEltll{n,j}\Delta\te C^\text{i}$
			\tcp*{Update inelastic deformation}
			
			$j \gets j +1$
		}
	\end{small}
\end{algorithm}

In order to solve the nonlinear tensor-valued equation following from the implicit exponential integrator~\eqref{eq:exp_mod}, we use the \emph{Newton-Raphson} scheme given in Alg.~\ref{alg:Newton-Raphson}. To initialize the scheme in each time step $n$, an explicit integrator is used, i.e., $\mEltl{n-1} \te H$, following from the last step's inelastic deformation $\mEltl{n-1} \te C^\text{i}$, is used instead of $\mElt{n} \te H$.
The iteration number $j$ a tensor belongs to is given as an index in the top left after the time increment number.
Within the iterative scheme, the \emph{Kelvin-Mandel notation} is used to represent symmetric 2nd order tensors as vectors, e.g., $\mEltll{n,j}\te R\in\Sym$ as $\mEltll{n,j}\vect R\in\R^6$,  and 4th order tensors with minor symmetry as matrices, e.g., $\mEltll{n,j}\tttte K\in\Ln_4$ as $\mEltll{n,j}\matr K\in\R^{6\times 6}$.\footnote{It is worth mentioning that the tangent $\mEltll{n,j}\tttte K = \partial_{\mEltll{n,j}\te C^\text{i}}{\mEltll{n,j}\te R}$ does not have the major symmetry, i.e., $\mEltll{n,j} K_{abcd} \ne \mEltll{n,j} K_{cdab}$.}

\section{Physics-augmented neural network model}
\label{sec:pann}

Based on the finite strain viscoelasticity theory presented in Sect.~\ref{sec:framework}, we introduce a \emph{physics-augmented neural network (PANN)} model, a related prediction mode to compute the stress for a given deformation-time series, and a suitable training method. 

\subsection{Model formulation}
\label{sec:model_formulation}

Following the concept of PANNs, as many constitutive conditions as possible should be fulfilled by construction \cite{Rosenkranz2024,Linden2023,Benady2024,Klein2023,Kalina2024,Dammass2025b}. We achieve this by only describing the potentials $\psi^\text{eq}(\I^\text{eq})$,  $\psi^\text{neq}(\I^\text{neq})$ and  $\phi^{*}(\I^{\phi^*})$ with suitable neural networks that guarantee the required properties, e.g., convexity w.r.t. to the thermodynamic forces to imply thermodynamic consistency. The overall structure of the model is then defined as in Sect.~\ref{sec:framework}, i.e., based on the concept of GSMs.

\subsubsection{Free energy}

We begin by formulating the free energy, which is decomposed additively according to Eq.~\eqref{eq:psi}, i.e., $\psi(\te F, \mte C^\text{i},\tilde p) = \psi^\text{eq}(\I^\text{eq}) + \psi^\text{neq}(\I^\text{neq}) + \psi^\text{inc}(J,\tilde p)$, where the last term remains unchanged.

\paragraph{Equilibrium part}

\begin{figure}
	\centering
	\includegraphics{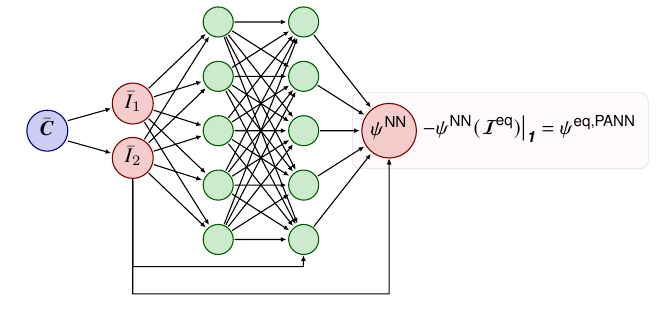}
	\caption{Neural network-based potential $\psi^\text{eq,PANN}$ for the description of the free energy equilibrium part  of the finite strain viscoelastic PANN. A monotonic FICNN with skip connections is used, where the network inputs are the invariants $\I^\text{eq} =(\bar I_1, \bar I_2)$ of the isochoric right Cauchy-Green deformation $\bte C$. The correction term $\psi^\text{NN}(\I^\text{eq})\big|_\one$ enforces zero energy in the undeformed state.}
	\label{fig:PANN_model_eq}
\end{figure}

For the equilibrium part we define the energy functional
\begin{align}
	\psi^\text{eq,PANN}(\I^\text{eq}) := \psi^\text{NN}(\I^\text{eq}) - \psi^\text{NN}(\I^\text{eq})\big|_\one \; ,
	\label{eq:PANN_eq}
\end{align}
with $\psi^\text{NN}(\I^\text{eq})$ being a \emph{monotonic} and \emph{fully input convex neural network (FICNN)}
\cite{Klein2021,Linden2023,Dammass2025b}. This network is constructed according to the FICNNs proposed by Amos~et~al.~\cite{Amos2017}, but with additional non-negativity constraints on the weights in the first hidden layer and the skip connections to enforce monotonicity. The weights and biases are collected in $\ve \theta^\text{eq} \in \mathscr{F\!i\!c\!n\!n}$, where the introduced set includes the non-negativity constraints on the weights \cite{Kalina2024,Dammass2025b}. The neural network-based representation of the equilibrium part is shown in Fig.~\ref{fig:PANN_model_eq}.

Since the equilibrium energy~\eqref{eq:PANN_eq} depends on the invariants $\I^\text{eq}$, it fulfills \emph{objectivity} and \emph{material symmetry}. 
As shown in \cite[Theorem~1]{Dammass2025a}, \emph{zero stress in the undeformed state}, i.e., 
$\te P^\text{eq,PANN}|_{\one} = \zero$,
is guaranteed since invariants of the isochoric part $\bte C\in\Sym\cap\SL$ are used.
Furthermore, the usage of the isochoric invariants and the correction term $ - \psi^\text{NN}(\I^\text{eq})\big|_\one$ enforce $\psi^\text{eq,PANN}(\I^\text{eq})|_{\one}=0$ and $\psi^\text{eq,PANN}(\I^\text{eq})\ge 0 \forall \te F \in \GLp$  by construction, cf. \cite[Theorem~3]{Dammass2025a}. 

Finally, due to the use of the monotonic FICNN, we ensure that the equilibrium energy is a \emph{polyconvex} functional of the argument $\te F$ in the sense of Ball~\cite{Ball1976}, cf. \cite{Klein2021,Linden2023,Dammass2025b,Dammass2025a}.\footnote{Note that the isochoric invariant $\bar I_2$ is not elliptic and thus not polyconvex in the case of compressible hyperelasticity \cite{Hartmann2003a}. However, for the special case of incompressible hyperelasticity $\bar I_2$ is elliptic \cite[Remark~2.1]{Klein2026}.}.
Note that this does not mean that the entire viscoelastic GSM model is polyconvex.
As we will show in Sect.~\ref{sec:PANN_lin}, the monotonic FICNN also guarantees a non-negative initial shear modulus $\mu\ge 0$ of the equilibrium part.

\paragraph{Non-equilibrium part}

\begin{figure}
	\centering
	\includegraphics{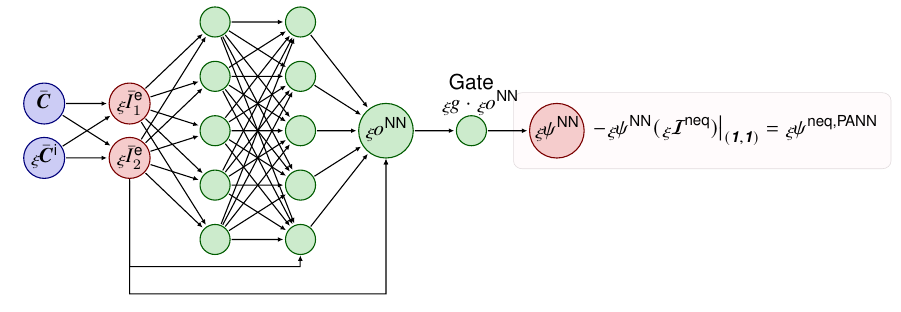}
	\caption{Neural network-based potential $\mEl \psi^\text{neq,PANN}$ for the description of the $\xi$th free energy non-equilibrium part of the finite strain viscoelastic PANN. A monotonic FICNN with skip connections is used, where the network inputs are the invariants $\mEl \I^\text{neq} = (\mEl \bar I_1^\text{e}, \mEl \bar I_2^\text{e})$ of the isochoric part of the $\xi$th elastic right Cauchy-Green deformation $\mEl \bte C^\text{e}$. A gate layer is placed behind the FICNN, which has the task of switching off unneeded Maxwell elements during training. The correction term $\mEl\psi^\text{NN}(\mEl\I^\text{neq})\big|_{(\one,\one)}$ enforces zero energy in the unloaded state.}
	\label{fig:PANN_model_neq}
\end{figure}

As discussed in Sect.~\ref{sec:framework}, our model represents a finite strain version of a \emph{generalized Maxwell model} with $N\in\N$ Maxwell elements. Thus, for the PANN, the additive decomposition into the energies of the individual Maxwell elements is also selected.	
Equivalently to the equilibrium part, we construct the non-equilibrium potentials based on FICNNs with invariant sets $\mEl \I^\text{neq}$ as input. For each Maxwell element, a tailored architecture consisting of a \emph{monotonic FICNN} and a \emph{trainable gate layer} is used, i.e., $\mEl\psi^\text{NN}: \R^2 \to \R_{\ge 0}\,,\, \mEl\I^\text{neq} \mapsto\mEl\psi^\text{NN}(\mEl\I^\text{neq}):= \left(\mEl \ell^\text{gate}\circ \mEl o^\text{NN}\right)(\mEl\I^\text{neq})$.
Weights and biases of the $N$ FICNNs are collected in $\ve \theta^\text{neq} \in \mathscr{F\!i\!c\!n\!n}$.
The task of the trainable gate layer is to remove unneeded Maxwell elements from the model during training. 
It is defined by 
\begin{align}
	\mEl \ell^\text{gate}: \R_{\ge 0} \to \R_{\ge 0}, \mEl o^\text{NN} \mapsto \mEl o^\text{NN}  \cdot \mEl g \text{ with }
	\mEl g := \min(1,\gamma\tanh(\epsilon \mEl \theta^\text{gate})) \in[0,1] \; ,
	\label{eq:gate_layer}
\end{align}
where $\gamma,\epsilon\in\R_{>0}$ are hyper parameters and $\mEl \theta^\text{gate}\in[0,1]$, $\xi\in\{1,2,\ldots,N\}$ are trainable variables. Thus, we have the additional set $\ve \theta^\text{gate} \in \mathscr{G\!a\!t\!e}:=\left\{ \ve \theta^\text{gate} \in \R^N \, | \, \mEl \theta^\text{gate} \in[0,1] \right\}$. The gate technique is adapted from \cite{Kalina2025}.

Equivalently to the equilibrium part~\eqref{eq:PANN_eq}, the entire non-equilibrium part is defined by
\begin{align}
	\psi^\text{neq,PANN}(\I^\text{neq}) := \sum_{\xi=1}^{N}
	\underbrace{\left(
		\mEl\psi^\text{NN}(\mEl\I^\text{neq}) - \mEl\psi^\text{NN}(\mEl\I^\text{neq})\big|_{(\one,\one)}
		\right)}_{\mEl \psi^\text{neq,PANN}(\mEl\I^\text{neq})} \; .
	\label{eq:PANN_neq}
\end{align}
The chosen architecture is depicted in Fig.~\ref{fig:PANN_model_neq}.
By using the ansatz~\eqref{eq:PANN_neq}, we also fulfill \emph{objectivity}, \emph{material symmetry} and \emph{invariance w.r.t. the rotational part of $\mEl \te F^\text{i}$} as well as, similar to the equilibrium part, ensure
\begin{align}
	\mEl\psi^\text{neq,PANN}(\mEl\I^\text{neq})|_{(\one,\one)} = 0 \; , \; 
	\mEl \te P^\text{neq,PANN}|_{(\one,\one)}=\zero \; , \; \mEl \te A^\text{neq,PANN}|_{(\one,\one)}=\zero 
\end{align}
for the undeformed state and 
$\mEl\psi^\text{neq,PANN}(\mEl\I^\text{neq}) \ge 0 \, \forall \te F\in\GLp, \mEl \te C^\text{i} \in \Sym \cap \GLp$.

Finally, due to the use of the \emph{monotonic FICNNs}, it is ensured that the non-equilibrium energies are \emph{polyconvex} functionals of the arguments $\mEl\te F^\text{e}$ in the sense of Ball~\cite{Ball1976}, cf. \cite[Sect.~5.1.1]{Gurses2007}. 
As for the equilibrium part, this also guarantees non-negative initial shear modules $\mEl\mu\ge 0$ of the Maxwell elements, cf. Sect.~\ref{sec:PANN_lin}.

\subsubsection{Dual dissipation potential}

\begin{figure}
	\centering
	\includegraphics{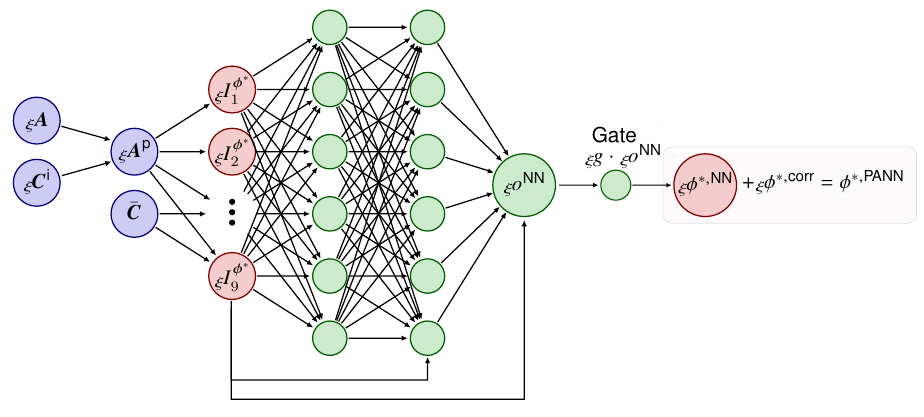}
	\caption{Neural network-based potential $\mEl \phi^{*,\text{PANN}}$ for the description of the $\xi$th dual dissipation potential of the finite strain viscoelastic PANN. A monotonic FICNN with skip connections is used, where the network inputs are mixed isotropic invariants $\mEl\I^{\phi^*} = (\mEl I_1^{\phi^*}, \mEl  I_2^{\phi^*},\ldots,\mEl  I_9^{\phi^*})$ of the $\xi$th projected thermodynamic forces $\mEl \te A^\text{p}$ and the isochoric right Cauchy-Green deformation $\bte C$. A gate layer is placed behind the FICNN, which has the task of switching off unneeded Maxwell elements during training. The correction terms $\mEl \phi^{*,\text{corr}}$, defined in Eq.~\eqref{eq:PANN_phi}, enforce $\mEl\phi^{*,\text{PANN}}(\mEl \I^{\phi^*})|_{(\mEl \te A^\text{p}(\zero,\mEl \te C^\text{i}),\bte C)}=0$ and $\partial_{\mEl\te A}\mEl\phi^{*,\text{PANN}}|_{(\mEl \te A^\text{p}(\zero,\mEl \te C^\text{i}),\bte C)} = \zero$.}
	\label{fig:PANN_model_diss}
\end{figure}

After the description of the free energy expressions above, we introduce a PANN approach for the dual dissipation potential. As discussed in Sect.~\ref{sec:framework}, we choose the specific structure given in  Eq.~\eqref{eq:mod_diss}, i.e., $\mEl\phi^*(\mEl\te A^\text{p},\bte C)$ with $\mEl\te A^\text{p}=\mEl\te A^\text{p}(\mEl\te A, \mEl \te C^\text{i})$ according to Eq.~\eqref{eq:projection}, to enforce the inelastic deformations to stay unimodular, i.e., $\mEl \te C^\text{i} \in \Sym\cap\SL$, during evolution, cf. Theorem~\ref{prop:unimodular}. To enforce \emph{objectivity}, \emph{material symmetry} and \emph{invariance w.r.t. the rotational part of $\mEl \te F^\text{i}$}, we choose the invariant sets $\mEl \I^{\phi^*}$ build from $\mEl \te A^\text{p}(\mEl\te A, \mEl\te C^\text{i})$ and $\bte C$ according to Eq.~\eqref{eq:invariants_phi}, that are convex w.r.t. $\mEl \te A$, cf. \ref{app:invariants_dissipation}. By applying an additive decomposition once again, we define
\begin{align}
	\phi^{*,\text{PANN}}(\I^{\phi^*}):= \sum_{\xi=1}^{N}
	\underbrace{
		\left(\mEl\phi^{*\text{,NN}}(\mEl\I^{\phi^*}) - \mEl\phi^{*\text{,NN}}(\mEl\I^{\phi^*})\big|_{(\mEl \te A^\text{p}(\zero,\mEl \te C^\text{i}),\bte C)}
		- \sum_{\alpha\in\{1,6,8\}}
		\diffp{\mEl\phi^{*\text{,NN}}}{\mEl I_\alpha^{\phi^*}}\Bigg|_{(\mEl \te A^\text{p}(\zero,\mEl \te C^\text{i}),\bte C)} \mEl I_\alpha^{\phi^*}
		\right)}_{\mEl\phi^{*\text{,PANN}}(\mEl\I^{\phi^*})} \; ,
	\label{eq:PANN_phi}
\end{align}
where the neural networks $\mEl\phi^{*\text{,NN}}(\mEl\I^{\phi^*})$ are \emph{monotonic FICNNs} combined with \emph{trainable gate layers} as already used for the non-equilibrium energies:
$\mEl\phi^{*,\text{NN}}: \R^9 \to \R_{\ge 0}\,,\, \mEl\I^{\phi^*} \mapsto\mEl\phi^{*,\text{NN}}(\mEl\I^{\phi^*}):= \left(\mEl \ell^\text{gate}\circ \mEl o^\text{NN}\right)(\mEl\I^{\phi^*})$.
The FICNNs' parameters are collected in $\ve \theta^{\phi^*} \in \mathscr{F\!i\!c\!n\!n}$. As the gates are shared with the non-equilibrium energies, no additional trainable variables enter here. The proposed NN-based potential is visualized in Fig.~\ref{fig:PANN_model_diss}.

With the chosen architecture we ensure that the
individual potentials $\mEl\phi^{*,\text{PANN}}(\mEl \I^{\phi^*})$ are convex and monotonic in $\mEl \I^{\phi^*}$ and thus convex in $\mEl\te A$.
With the second term we enforce $\mEl\phi^{*,\text{PANN}}(\mEl \I^{\phi^*})|_{(\mEl \te A^\text{p}(\zero,\mEl \te C^\text{i}),\bte C)}=0$ and with the last term we set the gradient for $\mEl \te A=\zero$ to 
\begin{align}
	\partial_{\mEl\te A}\mEl\phi^{*,\text{PANN}}|_{(\mEl \te A^\text{p}(\zero,\mEl \te C^\text{i}),\bte C)} = \zero \; \forall \mEl\te C^\text{i}, \bte C\in \Sym \; .
\end{align}
Note that the latter two properties in combination with the convexity imply 
$\mEl\phi^{*,\text{PANN}}(\mEl\I^{\phi^*})\ge 0 \; \forall \mEl\te A, \mEl\te C^\text{i}, \bte C\in \Sym$, cf. Footnote~\ref{foot:conditions_phi}. Also note that $(\mEl \te A = \zero) \Rightarrow (\mEl \te A^\text{p} = \zero)$ but  $(\mEl \te A^\text{p} = \zero) \nRightarrow (\mEl \te A = \zero)$.

\begin{rmk}
	It is worth noting that a formulation of the dual dissipation potentials based on \emph{partially input convex neural networks (PICNNs)} \cite{Amos2017} is also possible, see \cite{Rosenkranz2024}. Such an approach is more flexible but the number of trainable variables in the network increases. 
	Since the selected PANN model, which is based exclusively on FICNNs, has proven to be sufficiently flexible for the examples considered, we will not discuss PICNNs further here.
\end{rmk}

\subsubsection{Reduction to linear viscoelasticity at small strains}
\label{sec:PANN_lin}

As shown in Sect.~\ref{sec:linearization}, the proposed finite strain model can be simplified to the well-known \emph{linear viscoelasticity at small strains} by Taylor expansion of the potentials up to the quadratic order and subsequent linearization of the kinematic quantities. Since the selected NN approaches represent only a special case of the general model, this also applies to the PANN defined by the potentials~\eqref{eq:PANN_eq}, \eqref{eq:PANN_neq} and \eqref{eq:PANN_phi}.

To investigate the relation of the initial material constants $\mu^\text{PANN},\mEl\mu^\text{PANN},\mEl\eta^\text{PANN}$ with the NNs' weights, we consider the scalar-valued output of an arbitrary FICNN with input $\mte X \in \R^n$ and linear activations in the output that is defined by
\begin{align}
	g^\text{NN}(\mte X) = \sum_{\alpha=1}^{N^{\text{NN},H}} W_\alpha o_\alpha^{[H]}(\mte X) + \sum_{\beta=1}^{n} S_\beta X_\beta +B \in \R  \; ,
\end{align}
where $N^{\text{NN},H}\in\N$ is the number of neurons in the last hidden layer, $o_\alpha^{[H]}\in\R_{\ge 0}$ the $\alpha$th output of the last hidden layer and $W_\alpha, S_\beta\in \R_{\ge 0}$ the weights of the output layer and the skip connections to the output as well as $B\in \R$ the bias, respectively \cite{Kalina2024}. As can be seen from Eqs.~\eqref{eq:lin_equilibrium}, \eqref{eq:lin_nonequilibrium} and \eqref{eq:lin_dissipation}, the initial material parameters are related to the first derivative of the potentials w.r.t. to the invariants. Thus, we have to analyze the gradient
\begin{align}
	\diffp{g^\text{NN}}{X_\gamma} = 
	\sum_{\alpha=1}^{N^{\text{NN},H}} W_\alpha \diffp{o_\alpha^{[H]}(\mte X)}{X_\gamma} +  S_\gamma \; .
	\label{eq:grad_ficnn}
\end{align}
By using Eq.~\eqref{eq:grad_ficnn}, we find 
\begin{align}
	\mu^\text{PANN} &= 2\sum_{\gamma=1}^{2}\diffp{\psi^\text{NN}}{\bar I_\gamma}\Bigg|_{\one}
	= 2\sum_{\gamma=1}^{2} \left(
	\sum_{\alpha=1}^{N^{\text{NN},H}} W_\alpha \diffp{o_\alpha^{[H]}(\I^\text{eq})}{\bar I_\gamma}\Bigg|_{\one}
	+ S_\gamma
	\right) \ge 0 \\
	\mEl\mu^\text{PANN} &= 2\sum_{\gamma=1}^{2}\diffp{\mEl\psi^\text{NN}}{\mEl\bar I^\text{e}_\gamma}\Bigg|_{(\one,\one)} =
	2\sum_{\gamma=1}^{2} \left(
	\sum_{\alpha=1}^{N^{\text{NN},H}} \mEl W_\alpha \diffp{\mEl o_\alpha^{[H]}(\mEl\I^\text{neq})}{\mEl\bar I^\text{e}_\gamma}\Bigg|_{(\one,\one)}
	+\mEl S_\gamma
	\right) \ge 0 \\
	\mEl\eta^\text{PANN} &= \left[
	2\sum_{\gamma\in\{2,7,9\}}\diffp{\mEl\phi^{*,\text{NN}}}{\mEl\bar I^{\phi^*}_\gamma}\Bigg|_{(\zero,\one,\one)}
	\right]^{-1} 
	= \left[2\sum_{\gamma\in\{2,7,9\}} \left(
	\sum_{\alpha=1}^{N^{\text{NN},H}} \mEl W^*_\alpha \diffp{\mEl o_\alpha^{*,[H]}(\mEl\I^{\phi^*})}{\mEl\bar I^{\phi^*}_\gamma}\Bigg|_{(\zero,\one,\one)}
	+ \mEl S^*_\gamma
	\right)\right]^{-1} \ge 0
\end{align}
Thus, $\mu,\mEl \mu,\mEl\eta\ge 0$ are guaranteed due to the use of the \emph{monotonic FICNNs}. In addition, we find the useful relation that the initial shear modules depend linearly on the weights of the output layer and of the skip connections to the output, i.e., $k W_\alpha \wedge k S_\beta \Rightarrow k \mu^\text{PANN} \forall k\in\R_{\ge 0}$ and 
$k \mEl W_\alpha \wedge k \mEl S_\beta \Rightarrow k \mEl\mu^\text{PANN} \forall k\in\R_{\ge 0}$. Similarly, we find $k^{-1} \mEl W^*_\alpha \wedge k^{-1} \mEl S^*_\beta \Rightarrow k \mEl\eta^\text{PANN} \forall k\in\R_{\ge 0}$ 
for the initial viscosities. These relations will be very useful for the training described in Sect.~\ref{sec:training}.

\subsection{Prediction mode}

After formulating the model and analyzing the reduction to linear viscoelasticity, we will now consider how to \emph{predict stresses for a given load sequence}. We therefore assume that a trained model, given by the equilibrium energy $\psi^\text{eq,PANN}(\I^\text{eq})$, the non-equilibrium energy $\psi^\text{neq,PANN}(\I^\text{neq})$ and the dual dissipation potential $\phi^{*,\text{PANN}}(\I^{\phi^*})$, is already available and that the trainable parameters, collected in $\ve \theta\in\R^m$, are fixed. At this point, we would like to point out that the proposed viscoelastic PANN model does not differ fundamentally from a classical constitutive model, as only the potentials are replaced by neural networks. Thus, as our PANN is embedded into the framework proposed in Sect.~\ref{sec:framework}, the evolution of the internal variables $\mEl \te C^\text{i}$ is defined by Eq.~\eqref{eq:BiotEquation}.

To predict the stresses ${}^n\!\te P^\text{PANN}$, $n \in\incs:=\{1,2,\ldots,n_\text{inc}\}$ for a given load sequence $({}^n \Delta t, {}^n\!\te F)$, $n \in\incs$, one has to solve these $N$ evolution equations for the $N$ Maxwell elements in each time step to determine the inelastic deformations $\mElt{n} \te C^\text{i}$ from the implicit exponential integrator~\eqref{eq:exp_mod}. We solve these nonlinear equations with the \emph{Newton-Raphson scheme} given in Alg.~\ref{alg:Newton-Raphson}.
As initial conditions, we set ${}^0\!\te F= {}^0_\xi\!\te C^\text{i}=\one$.

To determine the full stress tensor, the pressure-like Lagrange multiplier $\tilde p$ has to determined from the boundary conditions. Within this work we use the \emph{plane stress} assumption. Thus, it holds 
\begin{align}
	[\te F] = 
	\begin{bmatrix}
		F_{11} & F_{12} & 0 \\
		F_{21} & F_{22} & 0 \\
		0 & 0 & F_{33} 
	\end{bmatrix} \text{ and }
	[\te P^\text{PANN}] = 
	\begin{bmatrix}
		P_{11}^\text{PANN} & P_{12}^\text{PANN} & 0 \\
		P_{21}^\text{PANN} & P_{22}^\text{PANN} & 0 \\
		0 & 0 & 0 
	\end{bmatrix} \; .
	\label{eq:plane_stress}
\end{align}
Due the incompressibility assumption, i.e., $J = 1$, we find from Eq.~\eqref{eq:plane_stress}${}_1$ that $F_{33} = (F_{11} F_{22} - F_{12}F_{21})^{-1}$. Furthermore, Eq.~\eqref{eq:plane_stress}${}_2$ in combination with Eq.~\eqref{eq:conditions}${}_1$ allows us to easily determine 
\begin{align}
	\tilde p^\text{PANN} = -F_ {33} (P^\text{eq,PANN}_{33} + \sum_{\xi=1}^{N}\mEl P^\text{neq,PANN}_{33})	
\end{align}
in a straightforward manner.

\subsection{Calibration of the model}
\label{sec:training}

In order to calibrate the model with experimental data, a suitable training method is required. Only variables that are experimentally accessible, e.g., from uniaxial tensile tests, are added to the data $\mathscr D:=\{\mathscr T_1,\mathscr T_2,\ldots,\mathscr T_{n_\text{load}}\}$ consisting of $n_\text{load}\in\N$ load case sets, each of the form  
\begin{align}
	\mathscr T_l:=\left\{({}^{l,1}\!\Delta t, {}^{l,2}\!\Delta t, \ldots, {}^{l,n_\text{inc}}\!\Delta t ),
	({}^{l,1}\!\te F, {}^{l,2}\!\te F, \ldots, {}^{l,n_\text{inc}}\!\te F ),
	({}^{l,1}\!\te P, {}^{l,2}\!\te P, \ldots, {}^{l,n_\text{inc}}\!\te P )
	\right\} \; .
	\label{eq:data_tuples}
\end{align}
In accordance with standard machine learning procedures \cite{Linden2023,Klein2021,Vlassis2020,Kollmannsberger2021}, we split the whole dataset $\mathscr D$ into \emph{calibration and test sets}, respectively: $\mathscr D = \mathscr D^\text{cal} \cup \mathscr D^\text{test}$ and $\varnothing = \mathscr D^\text{cal} \cap \mathscr D^\text{test}$. Thereby, the calibrated model should be able to generate reasonable predictions not only for the calibration but also for the test dataset which is crucial for \emph{generalizability}.
Since we consider a path dependent model, only entire load cases, collected in $\mathscr T_l$, are included in $\mathscr D^\text{cal}$ or $\mathscr D^\text{test}$, respectively. The indices $l$ of the calibration loadings are collected in the set $\cali$. For convenience, we summarize all trainable variables, namely weights and biases of the FICNNs as well as gate variables of the gate layers, in $\ve \theta = (\ve \theta^\text{eq}, \ve \theta^\text{neq},\ve \theta^{\phi^*}, \ve \theta^\text{gate})\in \mathscr{C\!o\!n\!s\!t}$. 

\begin{figure}
	\centering
	\includegraphics{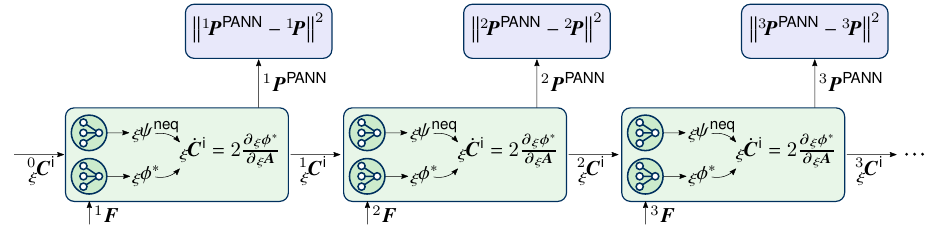}
	\caption{Schematic representation of the training process using the constrained optimization problem given in Eq.~\eqref{eq:optProb}. In each time step, the new internal variables $\mElt{n} \te C^\text{i}$ are obtained iteratively via the Newton-Raphson scheme given in Alg.~\ref{alg:Newton-Raphson}. Calculating the stress for time step $n$ thus requires the evaluation of all time steps $\{1,2,\ldots,n\}$ in advance. For simplicity's sake, only the case of a single load path is shown in the figure. The illustration is based on \cite{Rosenkranz2024}.}
	\label{fig:training}
\end{figure}

As can be seen from Eq.~\eqref{eq:data_tuples}, the internal variables $\mElt{n} \te C^\text{i}$ are not included into the data. However, in order to calculate the stress corresponding to a prescribed deformation time sequence, 
the knowledge of $\mElt{n} \te C^\text{i}$ is required. Thus, we solve the \emph{constrained optimization problem}
\begin{align}
	\begin{aligned}
		\hat{\ve \theta} &= \underset{\ve \theta\in \mathscr{C\!o\!n\!s\!t}}{\arg\min}\left(
		\frac{1}{n^{\te P}}\sum_{l\in\cali}
		\sum_{n=1}^{n_\text{inc}}\left\|\te P^\text{PANN}({}^{l,n}\!\te F, \mEltll{l,n}\te C^\text{i}(\ve \theta),\ve \theta) - {}^{l,n}\!\te P\right\|^2 + w^\text{gate}\mathscr L^\text{gate}(\ve \theta^\text{gate})\right) \\
		\;\; &\text{subject~to} \;\; 
		\mElt{n}\te C{}^\text{i} = \sqrt{\mEltl{n-1} \te C^\text{i}} \cdot \exp\left(\mElt{n} \hat{\te H} \, {}^n \!\Delta t\right) \cdot \sqrt{\mEltl{n-1} \te C^\text{i}}
		\; ,
	\end{aligned}
	\label{eq:optProb}
\end{align}
where $n^{\te P} := \frac{1}{3^2}\max\|{}^{l,n}\te P\|^2$, $l\in\cali, n\in\{1,2,\ldots,n_\text{inc}\}$.
This means, we have to solve the evolution equations within each iteration of the optimizer and differentiate through the Newton-Raphson scheme to get the parameter updates, cf. Fig.~\ref{fig:training} for a visualization.\footnote{Alternative training approaches for inelastic NN-based models are presented in \cite{Rosenkranz2024}. Instead of solving the evolution equations directly during training, the internal variables are provided by auxiliary FNNs or RNNs, and an additional loss term is added that penalizes deviation from the evolution equations. Although these methods allow for a significant speed-up of training in the case of implicit time discretization of the evolution equations, they are less accurate \cite{Rosenkranz2024}. The technique applied within this work is classified as \emph{integration method} in \cite{Rosenkranz2024}. To compute the gradients for the optimizer in a more efficient way, it is also possible to use the adjoint method \cite{Yan2025}.}
Since the stress $\te P^\text{PANN}$ in the first loss term, denoted as prediction loss $\mathscr L^\text{pred}$, is the gradient of the free energy w.r.t. $\te F$, this type of training is labeled as \emph{first order Sobolev training} \cite{Czarnecki2017,Vlassis2020,Kalina2025}. 
The additional loss term $\mathscr L^\text{gate}$, a penalty term based on the $p$-\emph{quasinorm} of the gates given by  
\begin{align}
	\mathscr L^\text{gate} := \frac{1}{n^\text{gate}}\left[\sum_{\xi=1}^N (\mEl g(\mEl\theta^\text{gate})+\delta)^p\right]^{\frac{1}{p}} 
	\text{ with } n^\text{gate}:= \left[N(1+\delta)^p\right]^{\frac{1}{p}} \; ,
	\label{eq:gate_loss}
\end{align}
enforces sparsity of the model w.r.t. the number of Maxwell elements and thus internal variables, cf. \cite{Flaschel2021,McCulloch2024,Kalina2025}. Thereby, $p \in \R_{> 0}$ and $N$ is the number of Maxwell elements. The parameter $\delta \ll 1$ prevents division by zero when differentiating. The weight $w^\text{gate}\in \R_{\ge 0}$ must be set appropriately in advance. All gates that fall below a value of $\num{1e-2}$ after training will be switched off.

We solve the optimization problem~\eqref{eq:optProb} with the \emph{Quasi-Newton optimizer SLSQP (sequential least squares programming)}. This allows for better results for small and moderately large networks than with stochastic gradient-based optimizers like Adam, cf. \cite[App.~G]{Kalina2024}, \cite[App.~E]{Otto2025}.
The implementation of the PANN model and the calibration workflow was realized using \emph{Python, TensorFlow} and \emph{SciPy}.

\begin{rmk}\label{remark:initial_adapt}
	Before starting the training, we modify the weights of the randomly initialized networks such that we get reasonable initial material parameters $\mu^\text{PANN}, \mEl \mu^\text{PANN}, \mEl \eta^\text{PANN}\ge 0$. This is done by using the results from the reduction to linear viscoelasticity, cf. Sect.~\ref{sec:PANN_lin}.
\end{rmk}

\begin{rmk}
	During calibration, the computation of the inelastic deformation tensors $\mEl \te C^\text{i}$ may become numerically unstable due to unfavorable values of the trainable parameters $\ve \theta$. Such situations can arise after parameter updates performed by the optimizer and may lead to a breakdown in the evaluation of $(\mEl \te C^\text{i})^{-1}$ when using TensorFlow’s built-in function \verb*|tf.linalg.inv|. To enhance the numerical robustness of the training, two modifications were introduced to avoid this in the implementation.
	
	First, the inverse is obtained by solving 
	\begin{align*}
		\mEl \te C^\text{i} \cdot (\mEl \te C^\text{i})^{-1} = \one \; ,
	\end{align*}
	i.e., three systems of linear equations have to  be solved to compute the inverse column-wise for each $\xi$. To this end, \verb*|tf.linalg.cholesky_solve| has been used.
	
	Second, as $\partial_{\mEl \te C^\text{i}}\det \mEl \te C^\text{i}=\det \mEl \te C^\text{i}(\mEl \te C^\text{i})^{-T}$, the determinant should also not computed directly via \verb*|tf.linalg.det| as this would lead to the calculation of the inverse via TensorFlow's in-build function during automatic differentiation again. Instead, the Cayley-Hamilton theorem is used to replace 
	\begin{align*}
		\det \mEl \te C^\text{i} = \frac{1}{3}\left(
		\tr (\mEl \te C^\text{i})^3 - \mEl I_1\tr (\mEl \te C^\text{i})^2 + \mEl I_2\tr \mEl \te C^\text{i}\right) 
		\; , \mEl I_1 = \tr\mEl \te C^\text{i}, \; \mEl I_2 = \frac{1}{2}\left(
		I_1^2 - \tr(\mEl \te C^\text{i})^2
		\right)
	\end{align*}
	with powers of $\mEl \te C^\text{i}$.
	
	Furthermore, solving the system of linear equations using the Newton-Raphson scheme according to Alg.~\ref{alg:Newton-Raphson} can lead to problems if the variables $\mEl \te C^\text{i}$ take on unfavorable values, even if the two stabilization techniques already described are applied. Thus, we start with a \emph{pre-training using an explicit exponential integrator}. After a few iterations, the weights are usually adjusted so that no further problems occur. Then the actual training (\emph{post-training}) with the implicit time integration method follows.
\end{rmk}

\section{Examples}
\label{sec:examples}

To illustrate the performance of the developed viscoelastic PANN, we will show calibration of the model using data from three examples. Thereby, \emph{interpolation behavior} of the PANN as well as the \emph{extrapolation behavior} is investigated.
All trainings were performed by applying the pre-training and post-training strategy as described in Sect.~\ref{sec:training}, where the SLSQP optimizer was used in both steps. 
Following \cite{Flaschel2021}, we have chosen $p=\frac{1}{4}$ for the exponent in the $p$-quasinorm. The parameters in the gate were chosen to $\gamma = 1.025$, $\epsilon = 2.5$ and $\delta = \num{1e-6}$, respectively \cite{Kalina2025}.
The value $w^\text{gate} = \num{5e-3}$ was found to be suitable and has been used in all training runs, see \ref{app:study_gate}. After pre-training, $w^\text{gate}$ was set to zero and all gates below a threshold of \num{1e-2} were deactivated.

In all examples, the PANN models were initialized with 5 Maxwell elements. Architectures with one hidden layer were used for all three NNs, with the networks for the energies having 8 neurons in the hidden layer and the network for the dual dissipation potential having 16 neurons.
Before training, the randomly initialized network parameters were modified such that $\mu^\text{PANN} = \mEl \mu^\text{PANN} = \mu^\text{av}$, with $\mu^\text{av}=1/6 \mu^\text{data}$ being the average initial shear modulus determined from initial slope of the calibration data. Afterwards, the parameters of the dual dissipation potential were modified such that $({}_1\tau^\text{PANN},{}_2\tau^\text{PANN},{}_3\tau^\text{PANN},{}_4\tau^\text{PANN},{}_5\tau^\text{PANN})=(5,10, 20, 40, 80)\,\text{s}$, with $\mEl \tau^\text{PANN}=\mEl \eta^\text{PANN}/\mEl \mu^\text{PANN}$ being the PANN's initial relaxation time, see Remark~\ref{remark:initial_adapt}.

All trainings were carried out with 8 CPUs each, whereby a high performance cluster (HPC) equipped with Intel Xeon Platinum 8470 CPUs was used. One training run takes about 15 to 20 minutes.

\subsection{Synthetic data}

Before we consider the case of real experimental data, let's first use synthetically generated data from a conventional model to evaluate the performance of the presented PANN approach.

\subsubsection{Conventional model as a ground truth}
As a ground truth, we use a model similar to the one presented in Rambausek~et~al.~\cite{Rambausek2022}, i.e., we  adapt it slightly so that it fits into the framework for incompressible finite strain viscoelasticity presented in Sect.~\ref{sec:framework}.\footnote{In contrast to \cite{Rambausek2022}, the Flory split is applied and the volumetric contributions are neglected within the free energy functionals for the equilibrium and non-equilibrium parts. The dual dissipation potential is used instead of the dissipation potential. The latter can be calculated by a Legendre Fenchel transformation, cf. Remark~\ref{rmk:biot_standard}. In addition, several Maxwell elements are used, whereas only one single element is used in \cite{Rambausek2022}.}

Within this model, the equilibrium and non-equilibrium contributions of the free energy are given by the \emph{neo-Hookean} potentials
\begin{align}
	\psi^\text{eq,gt}(\bar I_1) := \frac{\mu}{2} (\bar I_1 -3) \text{ and }
	\psi^\text{neq,gt}({}_1\!\bar I_1^\text{e},{}_2\!\bar I_1^\text{e},\ldots,{}_N\!\bar I_1^\text{e})
	:= 
	\sum_{\xi=1}^{N} \frac{\mEl\mu}{2} (\mEl \bar I_1^\text{e} -3) 
	\; ,\; \mu,\mEl \mu \in\R_{> 0}
	\; ,
	\label{eq:freeenergy_gt}
\end{align}
where an additive split of the non-equilibrium energy according to Eq.~\eqref{eq:psi} is applied. The dual dissipation potential is also additively decomposed and the contribution for the $\xi$th Maxwell element is defined as
\begin{align}
	\mEl \phi^{*,\text{gt}}(\mEl\tilde I_2^{\phi^*}) := \frac{1}{2\mEl \eta} \mEl\tilde I_2^{\phi^*}
	\; , \;
	\mEl\tilde I_2^{\phi^*} := \frac{1}{2} \tr\left(
	\mEl \tilde{\te A}{}^\text{p} \cdot \mEl \tilde{\te A}{}^\text{p}
	\right)
	\; ,\;
	\mEl \tilde{\te A}{}^\text{p} := \mEl \te A \cdot \mEl \te C^\text{i} - \frac{1}{3} \left(
	\mEl \te A : \mEl \te C^\text{i}
	\right) \one \in \Devi \; .
	\label{eq:dualdissipation_gt} 
\end{align}
From the potentials~\eqref{eq:freeenergy_gt} and \eqref{eq:dualdissipation_gt} with Eq.~\eqref{eq:BiotEquation}, one finds the specific form of the evolution equations for the ground truth model given by
\begin{align}
	\mEl\dot{\te C}{}^\text{i}
	= \frac{\mEl \mu}{\mEl \eta} \left(\te C - \frac{1}{3}\left((\mEl \te C^\text{i})^{-1}:\te C\right)\mEl \te C^\text{i}\right)
	\; ,
	\label{eq:evolution_gt}
\end{align} 
where it follows that $(\mEl \te C^\text{i})^{-1} : \mEl\dot{\te C}{}^\text{i}=0$ and thus $J^\text{i} = 1$ holds, cf. the incompressible case in \cite{Rambausek2022}.

To generate ground truth data for the calibration of our PANN model, we choose a model with three Maxwell elements and the material parameters given in Tab~\ref{tab:parameters_gt}.

\begin{rmk}
	The projected thermodynamic forces $\mEl \tilde{\te A}{}^\text{p}\in\Devi$ given in Eq.~\eqref{eq:dualdissipation_gt} are an alternative to $\mEl \te A^\text{p}\in\Sym$ as introduced in Eq.~\eqref{eq:projection}. Similar to $\mEl \te A^\text{p}$, formulating the dual dissipation potential in terms of $\mEl \tilde{\te A}{}^\text{p}$ enforces $(\mEl \te C^\text{i})^{-1} : \mEl\dot{\te C}{}^\text{i} = 0$, which implies unimodularity of $\mEl \te C^\text{i}$ during evolution, cf. Theorem~\ref{prop:unimodular}. 
	However, $\mEl \tilde{\te A}{}^\text{p}$ has the disadvantage that it is generally neither symmetric nor antimetric and therefore cannot be used to directly construct invariant sets using Boehler's method~\cite{Boehler1977}.
	It is also worth noting that the invariant $\mEl \tilde I_2^{\phi^*}$  can be represented by the set $\mEl\I^{\phi^*}\in\R^9$ given in Eq.~\eqref{eq:invariants_phi}, since $\mEl\I^{\phi^*}$ forms a functional basis of $(\mEl \te A^\text{p}, \bte C)$ and is thus a complete set.
\end{rmk}

\begin{rmk}
	It is worth mentioning that the chosen ground truth model~\eqref{eq:freeenergy_gt}--\eqref{eq:evolution_gt} coincides with the well-known model proposed by Reese~and~Govindjee~\cite{Reese1998} when it is specified for the incompressible case and neo-Hookean potentials are chosen for the free energy. This can be shown by transforming the evolution equation given in \cite{Reese1998} to Eq.~\eqref{eq:evolution_gt}. Another way to derive the model~\cite{Reese1998} using the GSM framework is described in \cite{Dammass2024}. 
\end{rmk}

\begin{table}
	\begin{center}
		\caption{Chosen parameters for the viscoelastic ground truth model according to Eqs.~\eqref{eq:freeenergy_gt} -- \eqref{eq:evolution_gt}. The relaxation times are defined as $\mEl \tau = \mEl \eta / \mEl \mu$.}
		\label{tab:parameters_gt}
		\begin{footnotesize}
			\begin{tabular}{lcccc}
				Part & Shear modulus $\mu/\text{MPa}$ & Shear modulus $\mEl \mu/\text{MPa}$ & Viscosity $\mEl \eta/\text{MPa}\cdot s$ & Relaxation time $\mEl \tau/\text{s}$\\
				\hline   
				Equilibrium & $0.3$ & $-$ &$-$ &$-$ \\
				Non-equilibrium $\xi=1$ & $-$ & $0.1$ & $0.5$ & $5.0$ \\
				Non-equilibrium $\xi=2$ & $-$ & $0.2$ & $4.0$ & $20.0$ \\
				Non-equilibrium $\xi=3$ & $-$ & $0.3$ & $24.0$ & $80.0$ 
			\end{tabular}
		\end{footnotesize}
	\end{center}
\end{table}

\subsubsection{Data generation}

\paragraph{Calibration data}
To mimic a real experimental setup, we use synthetic \emph{uniaxial} and \emph{equi-biaxial} tension tests for calibration. Following the works \cite{Asad2023,Rosenkranz2023,Rosenkranz2024}, we use \emph{smooth random walks}. These have the advantage that a wide variety of stretch rates and loading/unloading cases are included in each load path.
The stretch paths $\lambda(t)$ are created with cubic splines that connect a set of $n$ randomly sampled knots $({}^k\!\lambda^\text{knot}, {}^k t^\text{knot})\in\R_{>0}\times \R_{>0}$ with $k\in\{0,1,\ldots,n\}$ starting from ${}^0\!\lambda^\text{knot}=1$ and ${}^0t^\text{knot} = 0\,\text{s}$.
The time increments ${}^k\!\Delta t^\text{knot}$ are sampled from a uniform distribution: ${}^k\!\Delta t^\text{knot}\sim\mathcal U(\Delta t^\text{knot}_\text{min},\Delta t^\text{knot}_\text{max})$ with $\Delta t^\text{knot}_\text{min},\Delta t^\text{knot}_\text{max} \in \R_{>0}$.
The increments ${}^k\! \Delta \lambda^\text{knot}\sim\mathcal N(0,\sigma^2)$ are sampled from a normal distribution with mean zero and variance $\sigma^2$, where $\sigma = \Delta \lambda_\text{av}^\text{knot}/ \sqrt{2 / \pi}$ follows from the prescribed average stretch step width $\Delta \lambda_\text{av}^\text{knot}\in\R_{>0}$. If ${}^k\! \lambda^\text{knot} = {}^{k-1} \lambda^\text{knot} + {}^k\! \Delta \lambda^\text{knot}$ is not in $[\lambda_\text{min}^\text{knot},\lambda_\text{max}^\text{knot}]$, the increment is resampled. After sampling the knots, they are connected with cubic splines and divided into $n_\text{inc}$ time steps. The chosen hyperparameters and the resulting minimum and maximum stretches as well as absolute values of the stretch rates $|\dot \lambda|$ are given in Tab.~\ref{tab:parameters_training_lcs}.

\begin{table}
	\begin{center}
		\caption{Hyperparameters of the generated random walks for calibration and resulting maximum and minimum stretches in the loading direction(s) as well as absolute values of the stretch rates in the loading direction(s). For all random walks, the number of knots is $k = 20$.}
		\label{tab:parameters_training_lcs}
		\begin{footnotesize}
			\begin{tabular}{lccccccccc}
				Type & $\Delta \lambda_\text{av}^\text{knot}$ & $\lambda_\text{min}^\text{knot}$ &  $\lambda_\text{max}^\text{knot}$ & $\Delta t_\text{min}^\text{knot}/\text{s}$ & 
				$\Delta t_\text{max}^\text{knot}/\text{s}$ & $\min(\lambda)$ & $\max(\lambda)$ & $\min|\dot\lambda|/\text{s}^{-1}$ & $\max|\dot\lambda|/\text{s}^{-1}$\\
				\hline   
				Uniaxial & $0.1$ & $1.075$ &$2.0$ &$10.0$ &$50.0$ & $1.0$ &$1.92$ & \num{4.0e-6} &$0.024$\\
				Equi-biaxial & $0.05$ & $1.075$ & $1.5$ & $5.0$ &$25.0$ & $1.0$ &$1.46$ & \num{4.5e-5} &$0.021$\\
				Uniaxial & $0.1$ & $1.075$ & $2.0$ & $1.0$ &$5.0$ & $1.0$ &$1.92$ & \num{9.2e-6} &$0.157$
			\end{tabular}
		\end{footnotesize}
	\end{center}
\end{table}

\begin{table}
	\begin{center}
		\caption{Hyperparameters of the generated random walks for testing and resulting maximum and minimum stretches in the loading direction(s) as well as absolute values of the stretch rates in the loading direction(s). For the multiaxial loading, global maximum and minimum of both in-plane stretches $\lambda_1,\lambda_2$ are given. For all random walks, the number of knots is $k = 20$.}
		\label{tab:parameters_testing_lcs}
		\begin{footnotesize}
			\begin{tabular}{lccccccccc}
				Type & $\Delta \lambda_\text{av}^\text{knot}$ & $\lambda_\text{min}^\text{knot}$ &  $\lambda_\text{max}^\text{knot}$ & $\Delta t_\text{min}^\text{knot}/\text{s}$ & 
				$\Delta t_\text{max}^\text{knot}/\text{s}$ & $\min(\lambda)$ & $\max(\lambda)$ & $\min|\dot\lambda|/\text{s}^{-1}$ & $\max|\dot\lambda|/\text{s}^{-1}$\\
				\hline   
				Uniaxial & $0.1$ & $1.075$ &$2.0$ &$5.0$ &$25.0$ & $1.0$ &$1.92$ & \num{1.2e-5} &$0.034$\\
				Multiaxial & $0.1$ & $0.5$ & $1.5$ & $3.0$ &$15.0$ & $0.71$ &$1.46$ & \num{4.7e-5} &$0.13$
			\end{tabular}
		\end{footnotesize}
	\end{center}
\end{table}

\paragraph{Test data}In order to test the PANN, we generate additional load cases. To analyze the interpolation behavior, we use another uniaxial random walk as well as various relaxation tests in which the stretch is increased linearly and then held constant.
To test the extrapolation behavior, we perform uniaxial loading-unloading tests with increased maximum stretch as well as maximum stretch rate compared to the training regime. Finally, we generate a \emph{multiaxial smooth random walk}. To this end, two independent stretch paths $\lambda_1(t)$ and $\lambda_2(t)$ as well as a path $\varphi(t) \in[-\pi,\pi]$ are generated similar to the calibration data. By setting $\te R=\one$ in $\te F=\te R \cdot \te U$, the multiaxial deformation then follows to $\te F(t) = \te Q(\varphi(t)) \cdot \diag\left(\lambda_1(t),\lambda_2(t),1/\sqrt{\lambda_1(t)\lambda_2(t)}\right) \cdot \te Q^T(\varphi(t)) \in \Sym\cap \SL$, $\te Q \in \SO$. The chosen hyperparameters and the resulting minimum and maximum stretches for the test random walks as well as absolute values of the stretch rates $|\dot \lambda|$ are given in Tab.~\ref{tab:parameters_testing_lcs}. For the multiaxial loading, global maximum and minimum of both in-plane stretches $\lambda_1,\lambda_2$ are given.

\subsubsection{Performance of the PANN model}

\paragraph{Calibration} 

\begin{figure}
	\centering
	\includegraphics{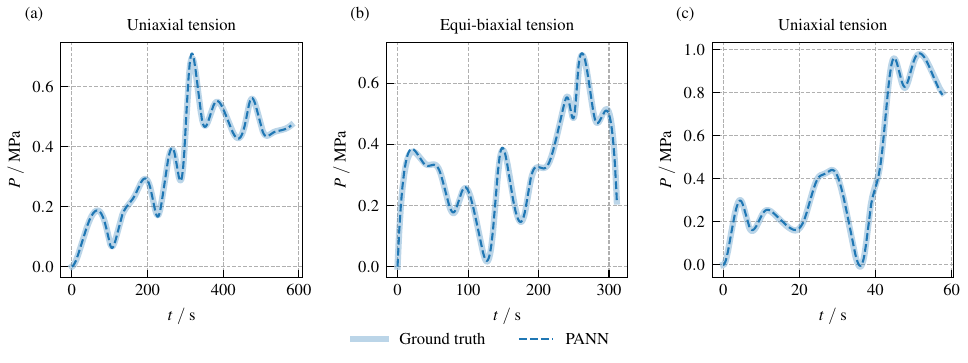}
	\caption{Stress responses of the trained PANN model compared to the ground truth model for the three calibration paths: (a) uniaxial random walk with $\max(\lambda) = 1.92$ and $\max|\dot\lambda|=0.024\,\text{s}^{-1}$, (b) equi-biaxial random walk with $\max(\lambda) = 1.46$ and $\max|\dot\lambda|=0.021\,\text{s}^{-1}$, and (c) uniaxial random walk with $\max(\lambda) = 1.92$ and $\max|\dot\lambda|=0.157\,\text{s}^{-1}$.}
	\label{fig:train_synth}
\end{figure}

During training, the number of active Maxwell elements was reduced from 5 to 2 through the application of the $\ell_p$ regularization. The comparison of ground truth and PANN predictions is shown in Fig.~\ref{fig:train_synth}. As can be seen, a very good approximation quality was achieved for all three training load cases.	

\begin{figure}
	\centering
	\includegraphics{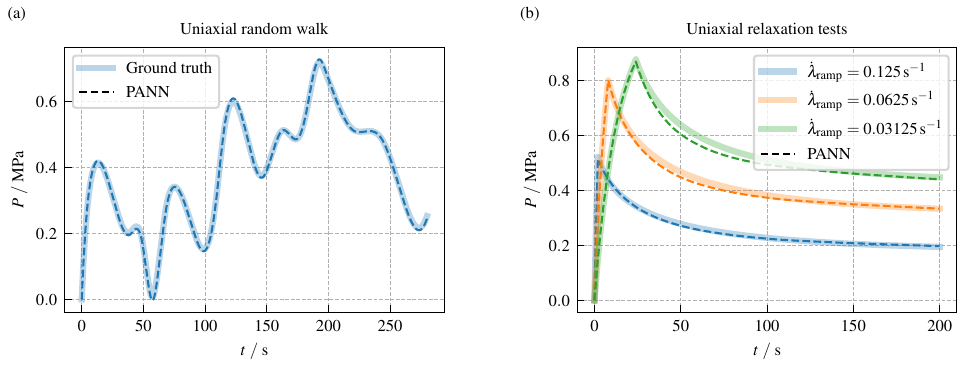}
	\caption{Stress responses of the trained PANN model compared to the ground truth model for two interpolation test scenarios: (a) uniaxial random walk with $\max(\lambda) = 1.92$ and $\max|\dot\lambda|=0.034\,\text{s}^{-1}$, and (b) uniaxial relaxation tests at different maximum stretches $\max(\lambda)\in\{1.25,1.5,1.75\}$ and stretch rates $\dot \lambda_\text{ramp}\in\{0.125,0.0625,0.03125\}\,\text{s}^{-1}$ during loading.}
	\label{fig:synth_test1}
\end{figure}

\paragraph{Test: Interpolation behavior}
As a first test load case, we consider the additional uniaxial random walk with similar minimum/maximum stretches and stretch rates as in the training case. The results are shown in Fig.~\ref{fig:synth_test1}(a). As for the training load cases, the quality of the prediction can be rated as very good. The investigated relaxation tests are shown in Fig.~\ref{fig:synth_test1}(b). Again, the PANN prediction corresponds well with the reference model. It should be noted that no load sequences involving long holding times with strain rates $\dot \lambda = 0 \,\text{s}^{-1}$ were included in the calibration data set. Nevertheless, due to its strong physical basis, the PANN is able to predict plausible behavior here.

\paragraph{Test: Extrapolation behavior}

\begin{figure}
	\centering
	\includegraphics{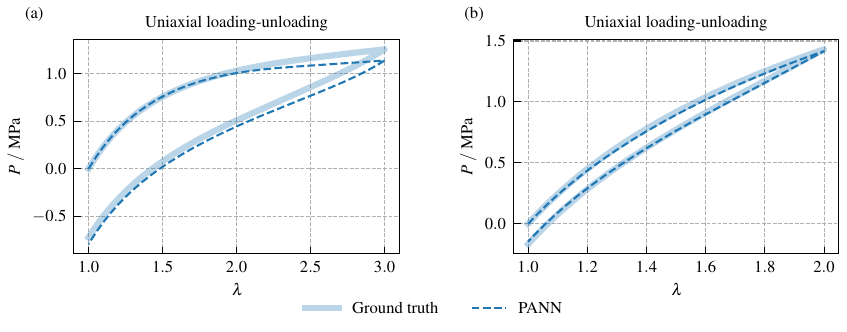}
	\caption{Stress responses of the trained PANN model compared to the ground truth model for two uniaxial loading-unloading test requiring extrapolation of the PANN: (a) $\max(\lambda) = 3$ and $\max|\dot\lambda|=0.04\,\text{s}^{-1}$, and (b) $\max(\lambda) = 2$ and $\max|\dot\lambda|=0.4\,\text{s}^{-1}$.}
	\label{fig:synth_test2}
\end{figure}

Since a reasonable constitutive model should provide plausible predictions for unseen loading paths, we also evaluate the extrapolation behavior of the PANN in addition to its interpolation behavior. An initial test involves \emph{uniaxial loading-unloading tests with increased maximum stretch or maximum stretch rate} compared to the training regime. The comparison between the reference and the predictions of the PANN is shown in Fig.~\ref{fig:synth_test2}. Here, too, a good agreement can be observed for the loadcase with increased stretch of $\max(\lambda) = 3$ with $\max|\dot\lambda|=0.04\,\text{s}^{-1}$. For the increased stretch rate of $\max|\dot\lambda|=0.4\,\text{s}^{-1}$ up to a stretch of  $\max(\lambda) = 2$ , the deviation to the ground truth is very low.

\begin{figure}
	\centering
	\includegraphics{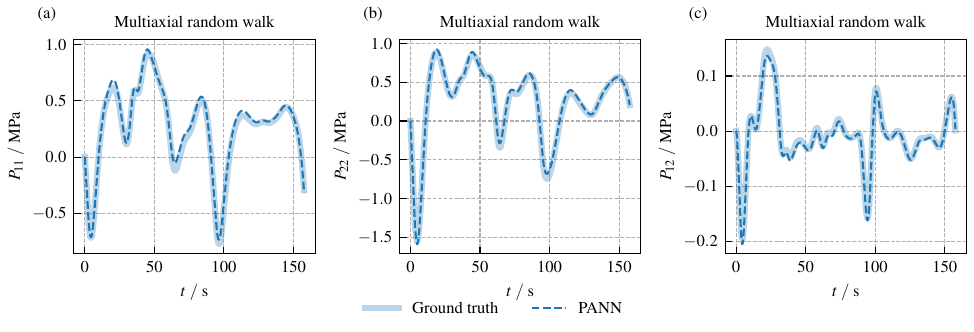}
	\caption{Stress responses of the trained PANN model compared to the ground truth model for a multiaxial random walk test requiring extrapolation of the PANN. Global maximum and minimum of both in-plane stretches $\max(\lambda_1,\lambda_2)=1.46$ and $\max(\lambda_1,\lambda_2)=0.71$. The maximum in-plane stretch rate is $\max(|\dot \lambda_1|,|\dot \lambda_2|)=0.13\,\text{s}^{-1}$.}
	\label{fig:synth_test3}
\end{figure}

Finally, the predicted in-plane stress components for the \emph{multiaxial smooth random walk} are shown in Fig.~\ref{fig:synth_test3}. Global maximum and minimum of both in-plane stretches are $\max(\lambda_1,\lambda_2)=1.46$ and $\min(\lambda_1,\lambda_2)=0.71$. The maximum in-plane stretch rate is $\max(|\dot \lambda_1|,|\dot \lambda_2|)=0.13\,\text{s}^{-1}$. Therefore, the PANN must not only extrapolate to multiaxial states, which differ from the uniaxial and equi-biaxial states observed in the calibration, but also extrapolate to the compression range. 
Here, too, the prediction quality of the PANN is very good.  This is particularly noteworthy considering that only uniaxial and equi-biaxial tests were used for calibration.	  
A similarly good extrapolation behavior has already been observed for elastic PANNs \cite{Linden2023} and viscoelastic PANNs in the small strain regime \cite{Rosenkranz2024}.

\subsection{Experimental data of VHB 4905 at $\vartheta = \SI{20}{\celsius}$ from Liao~et~al.~\cite{Liao2020}}

\begin{figure}
	\centering
	\includegraphics{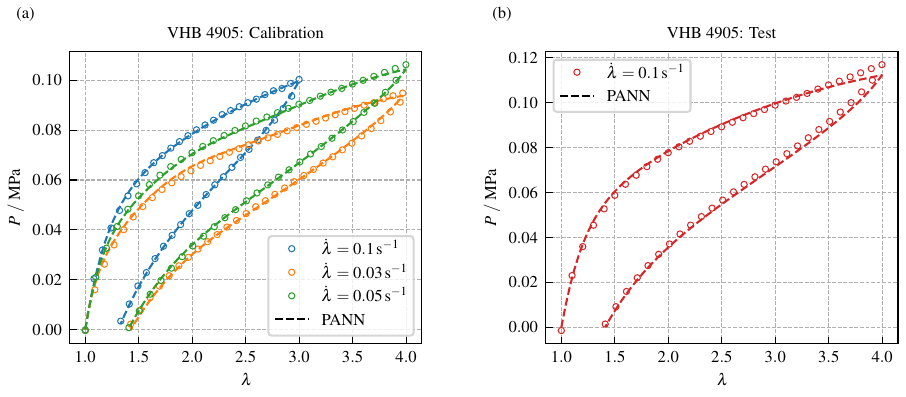}
	\caption{Results of the trained viscoelastic PANN for experimental uniaxial loading-unloading data of VHB 4905 at $\vartheta = \SI{20}{\celsius}$ from \cite{Liao2020}: (a) Calibration data and model prediction as well as (b) test data and model prediction.}
	\label{fig:vhb_4905}
\end{figure}

After testing the model with synthetically generated data, we now apply it to real experimental data. First, we consider \emph{uniaxial loading-unloading} tests of the polymer VHB 4905 at $\vartheta = \SI{20}{\celsius}$, taken from Liao~et~al.~\cite{Liao2020}. 
To control the time step size and ensure the same number of increments for all load cases, we interpolate between the measured points and  use the deformation time series obtained in this way for training.\footnote{\label{foot:inter}Piecewise cubic Hermite interpolating polynomials (PCHIPs) have been used for the interpolation of the loading and unloading, respectively. The total number of time steps was chosen to $n_\text{inc}=300$. The implementation was done via SciPy's \texttt{PchipInterpolator}.} We choose two load cases with $\lambda_\text{max} = 4$ and stretch rates $|\dot \lambda| \in\{0.03,0.05\}\,\text{s}^{-1}$ as well as one load case with $\lambda_\text{max} = 3$ and stretch rate $|\dot \lambda| = 0.1\,\text{s}^{-1}$ for calibration. The remaining load case with $\lambda_\text{max} = 4$ and $|\dot \lambda| = 0.1\,\text{s}^{-1}$ is used for testing.

The experimental data and predictions of the calibrated PANN are given in Fig.~\ref{fig:vhb_4905}. As with the synthetically generated data from the previous example, the model also achieves very good agreement with the real experimental data for the calibration load cases. During training, the number of active Maxwell elements was reduced from 5 to 2 through the application of the $\ell_p$ regularization. For the test load case that has not been considered for training, the prediction is still good, even though the model has to extrapolate here.

\subsection{Experimental data of VHB 4910 from Hossain~et~al.~\cite{Hossain2012}}

\begin{figure}
	\centering
	\includegraphics{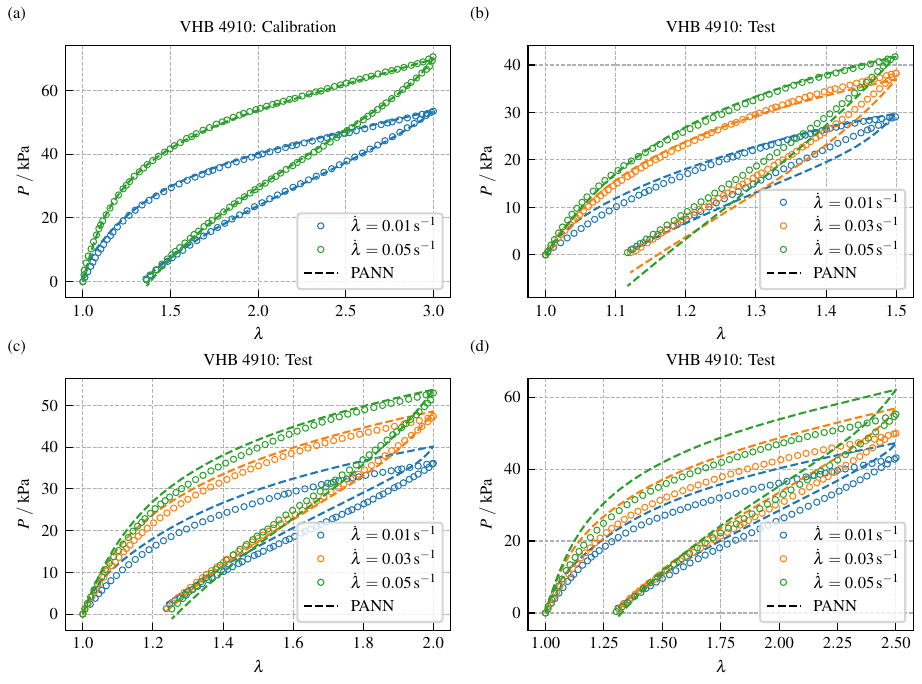}
	\caption{Results of the trained viscoelastic PANN for experimental uniaxial loading-unloading data of VHB 4910  from \cite{Hossain2012}: (a) Calibration data and model prediction for $\lambda_\text{max} = 3$ as well as (b) -- (c) test data and model prediction for different maximum stretches $\lambda_\text{max}\in\{1.5,2.0,2.5\}$.}
	\label{fig:vhb_4910}
\end{figure}

Within the last example, we use the experimental data of the polymer VHB 4910 from Hossain~et~al.~\cite{Hossain2012}.\footnote{Raw data was downloaded from \url{https://github.com/ConstitutiveANN/vCANN}.} As for VHB 4905, these experimental data contains \emph{uniaxial loading-unloading} tests at different maximum stretches and stretch rates. To control the time step size and ensure the same number of 300 increments for all load cases, we interpolate between the raw experimental data and use the deformation time series obtained in this way for training, cf. Footnote~\ref{foot:inter}.
As in Abdolazizi~et~al.~\cite{Abdolazizi2023a} and Holthusen~et~al.~\cite{Holthusen2024}, we choose two load cases with $\lambda_\text{max} = 3$ and stretch rates $|\dot \lambda| \in\{0.01,0.05\}\,\text{s}^{-1}$ for calibration. The remaining load cases are used for testing.

The experimental data and predictions of the calibrated PANN are given in Fig.~\ref{fig:vhb_4910}. The number of active Maxwell elements was automatically reduced from 5 to 2 through the application of the $\ell_p$ regularization during training. For this dataset a final number of 2 Maxwell elements is in line with \cite{Abdolazizi2023a}, where $\ell_1$ regularization was applied in a similar way. The prediction quality for calibration and test load cases is similar to that in \cite{Abdolazizi2023a,Holthusen2024}: The calibration data is very accurate. The test data for $\lambda_\text{max} \in \{1.5,2.0\}$ shows a fairly good match, whereas there are noticeable deviations for $\lambda_\text{max} = 2.5$. Similar observations have been made in the works \cite{Abdolazizi2023a,Holthusen2024}.

\begin{rmk}
	As also noted in Abdolazizi~et~al.~\cite[p.~13]{Abdolazizi2023a}, the noticeable deviations between the experimental data from Hossain~et~al.~\cite{Hossain2012} and the model predictions for the test case with maximum stretch $2.5$ in Fig.~\ref{fig:vhb_4910}(d) are likely due to experimental scatter. The loading paths at a fixed strain rate are non-identical, which suggest  a considerable uncertainty in parts of the experimental results. However, no information on the scatter of experimental data is provided in \cite{Hossain2012}. 
	Thus, it is not possible to get a perfect fit for both experiments, i.e., $\max(\lambda) = 3.0$ and $\max(\lambda) = 2.5$, at the same time.
\end{rmk}

\section{Conclusions}
\label{sec:conc}

In this work, a physics-augmented neural network approach for the data-driven modeling of finite strain incompressible viscoelasticity is proposed.
The formulation is embedded into the generalized standard materials framework and combines invariant-based neural network representations of the free energy and the dual dissipation potential with an implicit exponential integration scheme and automatic identification of the number of internal variables via $\ell_p$ regularization and trainable gates. The resulting model fulfills thermodynamic consistency and material symmetry by construction. In addition, the dual dissipation potential is constructed such that unimodularity of the inelastic deformations is guaranteed. The model shows excellent agreement with both synthetic and experimental data.

In summary, the presented viscoelastic PANN formulation represents a flexible material model that can serve as an alternative to classical models. 
The PANN is essentially not different from classical material models as only the functional descriptions of the potentials are replaced by neural networks.
Similar to conventional material models, the a priori incorporation of principles from constitutive modeling into PANNs ensures that the underlying physics is not violated even during extrapolation, thereby guaranteeing good generalization. This also allows comparatively small network architectures. The use of $\ell_p$ regularization enables the automatic elimination of unneeded Maxwell elements from the model.

Various applications and extensions of our approach are planned for the future. For example, an additional sparsification of the network as done in \cite{Fuhg2024a,McCulloch2024} is possible. 
Furthermore, the integration of the developed PANN model into Finite Element codes \cite{Kalina2023,Linka2021} or the calibration of the model via full-field data \cite{Linden2025a} and unsupervised learning \cite{Flaschel2023,Thakolkaran2022} are promising next steps. Finally, an extension to coupled problems \cite{Klein2024,Kalina2024,Zlatic2023} is possible.

\section*{Acknowledgment}
All presented computations were performed on a HPC-Cluster at the Center for Information Services and High Performance Computing (ZIH) at TU Dresden. The authors thus thank the ZIH for generous allocations of computer time. The authors thank the German Research Foundation (DFG) for the support within the Research Training Group GRK 2868 D${}^3$--Project Number 493401063. Finally, the authors want to thank Franz~Damma{\ss} and Brain~M.~Riemer for the fruitful discussions on the topic.

\section*{Usage of AI tools}
In preparing this work, the authors partially used ChatGPT, a generative AI tool, to improve the readability and language of the manuscript. After using this tool, the authors reviewed and revised the content as necessary and take full responsibility for the content of the published article.

\section*{CRediT authorship contribution statement}
\textbf{Karl A. Kalina:} Conceptualization, Formal analysis, Investigation, Methodology, Visualization, Software, Validation, Visualization, Writing -- original draft, Writing -- review \& editing, Funding acquisition. 
\textbf{J\"{o}rg Brummund:} Conceptualization, Formal analysis, Methodology, Writing – review \& editing. \textbf{Markus Kästner:} Resources, Writing -- review \& editing, Funding acquisition.

\appendix

\section{Convexity of the invariant set for the dual dissipation potential}
\label{app:invariants_dissipation}

In this appendix, we prove the convexity of the proposed invariant set $\mEl\I^{\phi^*}\in\R^9$ according to Eq.~\eqref{eq:invariants_phi} w.r.t. the thermodynamic forces $\mEl \te A$. This invariant set is used to 	
formulate the modified dual dissipation potential~\eqref{eq:mod_diss}.

\subsection{Convexity of the projection operation}

\begin{proposition}\label{prop:projection_convex}
	Let $\te X^\text{p} = \tttte L : \te X$, $\te X \in \Ln_2, \tttte L\in \Ln_4$, with $\tttte L=\text{const.}$, be a linear transformation of $\te X$ and $f: \Ln_2 \to \R, 
	\te X^\text{p} \mapsto f(\te X^\text{p})$ a functional that is convex w.r.t. $\te X^\text{p}$. Then $f(\te X^\text{p}(\te X))$ is convex w.r.t. $\te X$. 
\end{proposition}

\begin{proof}
	We analyze the Hessian of $f(\te X^\text{p}(\te X))$ w.r.t. $\te X$.
	By using the chain rule and accounting for the convexity of $f(\te X^\text{p})$ w.r.t. $\te X^\text{p}$, we find
	\begin{align}
		\delta \te X : \diffp{{}^2 f}{\te X\partial \te X} : \delta \te X = 
		(\tttte L : \delta \te X) : \diffp{{}^2 f}{\te X^\text{p}\partial \te X^\text{p}} : (\tttte L : \delta \te X) = \delta \te X^\text{p} : \diffp{{}^2 f}{\te X^\text{p}\partial \te X^\text{p}} : \delta \te X^\text{p} \ge 0 \; \forall \te X^\text{p}, \delta \te X^\text{p} \in \Ln_2 \; ,
	\end{align}
	where $\delta \te X^\text{p} = \tttte L : \delta \te X$.
\end{proof}

From Proposition~\ref{prop:projection_convex}, we find that convexity of the dual dissipation potentials $\mEl \phi^*(\mEl \te A^\text{p}(\mEl \te A,\mEl \te C^\text{i}), \bte C)$  w.r.t. $\mEl \te A^\text{p}$ implies convexity w.r.t. $\mEl \te A$.\footnote{Note that the projectors $\mEl\tttte P := \ttttes 1^\text{s} - 1/3 (\mEl \te C^\text{i})^{-1} \otimes \mEl \te C^\text{i}$ in $\mEl \te A^\text{p} := \mEl\tttte P:\mEl \te A$ are functions of $\mEl\te C^\text{i}$. Nevertheless, because $\mEl \te A$ are treated as independent constitutive variables, the projectors are still linear in $\mEl \te A$.} 
It is thus sufficient to prove convexity of  $\mEl\I^{\phi^*}$ w.r.t. $\mEl \te A^\text{p}$.

\subsection{Convexity of the invariants}

As shown in Rosenkranz~et~al.~\cite{Rosenkranz2024}, the invariants $\mEl I_1^{\phi^*} = \tr \mEl \te A^\text{p}$, 
$\mEl I_2^{\phi^*} = \frac{1}{2}\tr  \left(\mEl\te A^\text{p}\right)^2$, $\mEl I_3^{\phi^*} = \frac{1}{4}\tr  \left(\mEl\te A^\text{p}\right)^4$ are convex w.r.t. $\mEl \te A^\text{p}$. 

Thus, we only have to show the convexity of the mixed invariants
$\mEl I_6^{\phi^*}$, $\mEl I_7^{\phi^*}$, $\mEl I_8^{\phi^*}$ and $\mEl I_9^{\phi^*}$ in the following.
To prove this, we make use of the spectral decomposition. 

Consider the spectral decompositions 
\begin{align}
	\te S = \sum_{\alpha=1}^{N} S_\alpha \te M_\alpha \in \Sym \text{ and }
	\tilde{\te S} = \sum_{\beta=1}^{\tilde N} \tilde S_\beta \tilde{\te M}_\beta \in \Sym
	\label{eq:projection_tensors}
\end{align}
of two symmetric and positive semi-definite 2nd order tensors $\te S$ and $\tilde{\te S}$, with $S_\alpha, \tilde S_\beta\in\R_{\ge 0}$ being the eigenvalues, $\te M_\alpha, \tilde{\te M}_\beta\in\Sym$ the projection tensors and $N,\tilde N\in\{1,2,3\}$ the number of non-equal eigenvalues. The projection tensors $\te M_\alpha$ can be expressed via the eigenvectors $\ve N_\alpha \in \Ln_1$ with $|\ve N_\alpha|=1$ as 
\begin{align}
	\te M_\alpha &= \ve N_\alpha \otimes \ve N_\alpha \;,\; \alpha \in\{1,2,3\} \; \text{for } N=3 \; ,\\
	\te M_1 &= \ve N_1 \otimes \ve N_1 \; ,\; \te M_2 =  \one - \ve N_1 \otimes \ve N_1 \; \text{for } N=2 \; ,\\
	\te M_1 &= \one \; \text{for } N=1
\end{align}
and $\tilde{\te M}_\beta$ likewise \cite[Sect.~4.6]{Sahraee2023}.

\begin{lemma}\label{lemma:projection_tensors}
	Let $\te M_\alpha \in \Sym$, $\alpha\in\{1,\ldots,N\}$ and $\tilde{\te M}_\beta \in \Sym$, $\beta\in\{1,\ldots,\tilde N\}$ the projection tensors of two symmetric 2nd order tensors as introduced in Eq.~\eqref{eq:projection_tensors}, with $N,\tilde N\in\{1,2,3\}$ non-equal eigenvalues, respectively. Then it holds 
	\begin{align}
		\te M_\alpha : \tilde{\te M}_\beta \ge 0  \; .
	\end{align}
\end{lemma}

\begin{proof}
	The projection tensor(s) $\te M_\alpha$ can be expressed via the eigenvectors $\ve N_\alpha \in \Ln_1$ with $|\ve N_\alpha|=1$ as 
	\begin{align}
		\te M_\alpha &= \ve N_\alpha \otimes \ve N_\alpha \;,\; \alpha \in\{1,2,3\} \; \text{for } N=3 \; ,\\
		\te M_1 &= \ve N_1 \otimes \ve N_1 \; ,\; \te M_2 =  \one - \ve N_1 \otimes \ve N_1 \; \text{for } N=2 \; ,\\
		\te M_1 &= \one \; \text{for } N=1
	\end{align}
	and $\tilde{\te M}_\beta$ likewise.    	
	With $(\ve N_\alpha \cdot \tilde{\ve N}_\beta)^2\in[0,1]$ and $|\ve N_\alpha|=1$ we only get the non-negative products
	\begin{align}
		\te M_\alpha : \tilde{\te M}_\beta &= (\ve N_\alpha \cdot \tilde{\ve N}_\beta)^2 \ge 0 \; \forall \alpha,\beta\in\{1,2,3\} \; \text{if } N=3, \tilde N=3 \; ,\\
		\te M_\alpha : \tilde{\te M}_\beta &\ge |\ve N_\alpha|^2 - (\ve N_\alpha \cdot \tilde{\ve N}_1)^2 \ge 0 \; \forall \alpha\in\{1,2,3\}, \beta\in\{1,2\} \; \text{if } N=3, \tilde N=2\; ,\\
		\te M_\alpha : \tilde{\te M}_\beta &= |\ve N_\alpha|^2=1> 0 \; \forall \alpha\in\{1,2,3\}, \beta\in\{1\}
		\; \text{if } N=3, \tilde N=1\; ,\\
		\te M_\alpha : \tilde{\te M}_\beta &\ge |\ve N_1|^2 - (\ve N_1 \cdot \tilde{\ve N}_1)^2 \ge 0 \; \forall \alpha,\beta\in\{1,2\} \; \text{if } N=2, \tilde N=2\; ,\\
		\te M_\alpha : \tilde{\te M}_\beta &= 3- |\ve N_1|^2 \ge 0 \; \forall \alpha\in\{1,2\},\beta\in\{1\} \; \text{if } N=2, \tilde N=1\; ,\\
		\te M_\alpha : \tilde{\te M}_\beta &= 3 \ge 0 \; \forall \alpha,\beta\in\{1\} \; \text{if } N=1, \tilde N=1 \; .
	\end{align}
	The remaining three combinations are trivial. 
\end{proof}    

\begin{proposition}
	The mixed invariants $\mEl I_6^{\phi^*} = \tr  \left(\mEl \te A^\text{p}\cdot \bte C\right)$,
	$\mEl I_7^{\phi^*} = \frac{1}{2}\tr  \left((\mEl \te A^\text{p})^2\cdot \bte C\right)$,
	$\mEl I_8^{\phi^*} = \tr  \left(\mEl \te A^\text{p}\cdot \bte C^2\right)$ 
	and $\mEl I_9^{\phi^*} = \frac{1}{2}\tr  \left((\mEl \te A^\text{p})^2\cdot \bte C^2\right)$ are convex w.r.t. $\mEl \te A^\text{p}$.
\end{proposition}

\begin{proof}
	The mixed invariants $\mEl I_6^{\phi^*} = \tr  \left(\mEl \te A^\text{p}\cdot \bte C\right)$ and $\mEl I_8^{\phi^*} = \tr  \left(\mEl \te A^\text{p}\cdot \bte C^2\right)$ are linear in $\mEl \te A^\text{p}$ and thus convexity with respect to $\mEl \te A^\text{p}$ follows trivially, since the Hessian is zero.
	
	By analyzing the convexity condition for the Hessian of $\mEl I_7^{\phi^*}$ and using the spectral decompositions of $(\mEl\delta  \te A^\text{p})^2$ and $\bte C$, we find
	\begin{align}
		\mEl\delta  \te A^\text{p} : \diffp{{}^2\mEl I_7^{\phi^*}}{\mEl \te A^\text{p}\partial \mEl \te A^\text{p}} : \mEl\delta \te A^\text{p} 
		= \left(\mEl\delta  \te A^\text{p} \cdot \mEl\delta \te A^\text{p}\right) : \bte C = \sum_{\alpha=1}^{N_{\delta\te A^\text{p}}}\sum_{\beta=1}^{N_{\bte C}} (\mEl \delta A_\alpha^\text{p})^2 \bar \lambda_\beta^2 \te M^{\delta \te A^\text{p}}_\alpha : \te M^{\bte C}_\beta \ge 0 \; \forall \mEl \delta \te A^\text{p} \in \Sym \; .
	\end{align}
	By using Lemma~\ref{lemma:projection_tensors}, we  get 
	\begin{align}
		\mEl\delta  \te A^\text{p} : \diffp{{}^2\mEl I_7^{\phi^*}}{\mEl \te A^\text{p}\partial \mEl \te A^\text{p}} : \mEl\delta \te A^\text{p} 
		= \sum_{\alpha=1}^{N_{\delta\te A^\text{p}}}\sum_{\beta=1}^{N_{\bte C}} (\mEl \delta A_\alpha^\text{p})^2 \bar \lambda_\beta^2 \te M^{\delta \te A^\text{p}}_\alpha : \te M^{\bte C}_\beta \ge 0 \; \forall \mEl \delta \te A^\text{p} \in \Sym, \; \bte C\in\Sym\cap\SL \; .
	\end{align}
	Since, similar to $\bte C$, $\bte C^2$ is symmetric and positive definite,  the argumentation for $\mEl I_9^{\phi^*}$ is analogue to $\mEl I_7^{\phi^*}$.
\end{proof}

\section{Properties of the exponential integrators}
\label{app:exp_integrators}

Within this appended section, we discuss the properties of the exponential integrators~\eqref{eq:exp_orig} and \eqref{eq:exp_mod} for the numerical solution of the evolution equations~\eqref{eq:BiotEquation}. Thereby we make use of the well-known definition \cite[App.~B.1]{deSouzaNeto2008a}
\begin{align}
	\exp: \Ln_2 \to \Ln_2: \te X \mapsto
	\exp\left(\te X\right) := \sum_{k=0}^\infty  \frac{\te X^{k}}{k!} \; ,
	\label{eq:tensor_exponential}
\end{align}
and the properties $\exp(\te X) \cdot \te X = \te X \cdot \exp(\te X)$ as well as  $\det[\exp(\te X)] = 1$ if $\te X \in \Devi$ \cite[App.~B.1.1]{deSouzaNeto2008a}.

\subsection{Standard exponential map integrator}

We start with the "original" exponential map integrator from Eq.~\eqref{eq:exp_orig}.

\begin{theorem}\label{theorem:exact_exp_map_orig}
	Consider the ODE $\mEl\dot{\te C}{}^\text{i} = \mEl \te H \cdot \mEl \te C^\text{i}$ with $ \mEl \te H=\text{const.}$, 
	then the exponential integrator $\mEl\te C{}^\text{i} = \exp\left(\mEl \te H \, \Delta t \right) \cdot \mElt{0} \te C^\text{i}$, with $\Delta t= t-{}^0 t$ and $t\ge {}^0 t$, is an exact solution of the ODE with the initial condition $\te C^\text{i}(t={}^0 t) = \mElt{0} \te C^\text{i}$.
\end{theorem}

\begin{proof}
	Forming the time derivative of the exponential integrator for $\mEl \te H=\text{const.}$ and using Eq.~\eqref{eq:tensor_exponential} gives
	\begin{align}
		\mEl\dot{\te C}{}^\text{i} = \mEl \te H \cdot \left(\sum_{k=1}^\infty  \mEl \te H^{k-1} \frac{\Delta t^{k-1}}{(k-1)!}\right)  \cdot \mElt{0} \te C^\text{i}
		= \mEl \te H \cdot \left( \sum_{k=0}^\infty  \mEl \te H^{k} \frac{\Delta t^{k}}{k!} \right)  \cdot \mElt{0} \te C^\text{i}
		= 
		\mEl \te H \cdot \exp\left(\mEl \te H \, \Delta t\right) \cdot \mElt{0} \te C^\text{i}
		= \mEl \te H \cdot \mEl \te C^\text{i} \; . \label{eq:proof_solution}
	\end{align}
\end{proof}

\begin{theorem}\label{theorem:sym_exp_map_orig}
	Let $\mEltl{n-1} \te C^\text{i}\in\Sym\cap\SL$ the inelastic deformation of the last time step, i.e., $\mEltl{n-1} \te C^\text{i} = (\mEltl{n-1} \te C^\text{i})^T$ and $\det \mEltl{n-1} \te C^\text{i} = 1$, and $\mElt{n} \te H \in \Devi$, i.e., $\tr \mElt{n} \te H = 0$.
	Then the implicit exponential integrator 
	$\mElt{n}\te C{}^\text{i} = \exp\left(\mElt{n} \te H \, {}^n \!\Delta t\right) \cdot \mEltl{n-1} \te C^\text{i}$ 
	guarantees that the solution $\mElt{n}\te C{}^\text{i}$ is symmetric and unimodular.
\end{theorem}

\begin{proof}
	Unimodularity of the solution $\mElt{n}\te C{}^\text{i}$ directly follows from $\det\left[\exp\left(\mElt{n} \te H \, {}^n \!\Delta t\right)\right] = 1$ as $\tr \mElt{n} \te H = 0$.
	By using the definition of $\mElt{n} \te H$ from Eq.~\eqref{eq:defH}, it follows $\mElt{n} \te H = \mElt{n} \te B \cdot (\mElt{n} \te C^\text{i})^{-1}$ with $\mElt{n} \te B:=2 \partial_{\mElt{n}\te A}{\mEl\phi^*}$. From that we find 
	by inversion of Eq.~\eqref{eq:exp_orig} and with Eq.~\eqref{eq:tensor_exponential} that 
	\begin{align}
		(\mEltl{n-1} \te C^\text{i})^{-1} &=
		(\mElt{n} \te C^\text{i})^{-1} \cdot \exp\left(\mElt{n} \te B \cdot (\mElt{n} \te C^\text{i})^{-1} {}^n \!\Delta t\right)\\
		&=(\mElt{n} \te C^\text{i})^{-1} \cdot \left(
		\sum_{k=0}^\infty \left[\mElt{n} \te B \cdot (\mElt{n} \te C^\text{i})^{-1}\right]^k \frac{{}^n \!\Delta t^k}{k!}\right)
		\; .
		\label{eq:sym_proof}
	\end{align}
	As $\mEltl{n-1} \te C^\text{i}$ is symmetric, the right side of Eq.~\eqref{eq:sym_proof}, given by
	$(\mElt{n} \te C^\text{i})^{-1} + (\mElt{n} \te C^\text{i})^{-1} \cdot \mElt{n} \te B \cdot (\mElt{n} \te C^\text{i})^{-1} {}^n \!\Delta t + \ldots $,
	must also symmetric. Since the power series in ${}^n \!\Delta t$ must be symmetric for all  ${}^n \!\Delta t \in \R_{\ge 0}$, each tensor valued coefficient must be symmetric and it directly follows $\mElt{n} \te C^\text{i} \in \Sym$ which implies $\mElt{n} \te B\in\Sym$. Thus, the solution of the implicit exponential map is $\mElt{n} \te C^\text{i} \in \Sym \cap \SL$.   	
\end{proof}

\begin{rmk}
	Note that the symmetry and unimodularity of the inelastic deformations $\mElt{n} \te C^\text{i}$ from time step $n$ is only guaranteed for the solution of the nonlinear equation~\eqref{eq:exp_orig}. These properties do not hold for the intermediate results of the internal variables during solution by a Newton-Raphson  scheme. 
\end{rmk}

\subsection{Modified exponential map integrator}

After considering the "original" exponential map integrator, we will now analyze the modified version according to Eq.~\eqref{eq:exp_mod}.

\begin{theorem}\label{theorem:exact_exp_map_mod}
	Consider the ODE $\mEl\dot{\te C}{}^\text{i} = \mEl \te H \cdot \mEl \te C^\text{i}$ with $\mEl \te H=\text{const.}$
	and the initial condition $\te C^\text{i}(t={}^0 t) = \mElt{0} \te C^\text{i} \in \Sym$, 
	then the exponential integrator $\mEl \te C{}^\text{i} = \sqrt{\mElt{0} \te C^\text{i}} \cdot \exp\left(\mEl \hat{\te H} \, \Delta t\right) \cdot \sqrt{\mElt{0} \te C^\text{i}}$  with $\mEl\hat{\te H}:=\sym\left(\sqrt{(\mElt{0} \te C^\text{i})^{-1}} \cdot \mEl \te H \cdot \sqrt{\mElt{0} \te C^\text{i}}\right)$ and $\Delta t= t-{}^0 t$, $t\ge {}^0 t$ is an exact solution of the ODE.
\end{theorem}

\begin{proof}
	By applying the same technique as in Eq.~\eqref{eq:proof_solution}, we find
	\begin{align}
		\mEl\dot{\te C}{}^\text{i} =  \sqrt{\mElt{0} \te C^\text{i}} \cdot \mEl \hat{\te H} \cdot  \exp\left(\mEl \hat{\te H} \, \Delta t\right) \cdot \sqrt{\mElt{0} \te C^\text{i}}
	\end{align}
	for the time derivative of the exponential integrator for $\mEl \te H=\text{const}$. With the definition of $\mEl \hat{\te H}$ and by using $\mEl {\te H} \cdot \mElt{0} \te C^\text{i} = \mElt{0} \te C^\text{i} \cdot \mEl {\te H}{}^T$, which follows for $\Delta t=0$ from $\mElt{0}\dot{\te C}{}^\text{i} = \mEl \te H \cdot \mElt{0} \te C^\text{i}$, we get  
	\begin{align}
		\mEl\dot{\te C}{}^\text{i} = \frac{1}{2}\left(
		\mEl \te H \cdot \mEl {\te C}{}^\text{i} + \mElt{0} \te C^{i} \cdot \mEl \te H^T \cdot (\mElt{0} \te C^{i})^{-1} \cdot \mEl{\te C}{}^\text{i}
		\right) =
		\mEl \te H \cdot \mEl{\te C}{}^\text{i} \; .
	\end{align}
\end{proof}

\begin{lemma}
	\label{lemma:H_deviator}
	Let $\mElt{n}\hat{\te H}$ be defined according to Eq.~\eqref{eq:exp_mod}${}_2$ by $\mElt{n}\hat{\te H}:=\sym\left(\sqrt{(\mEltl{n-1} \te C^\text{i})^{-1}} \cdot \mElt{n} \te H \cdot \sqrt{\mEltl{n-1} \te C^\text{i}}\right)$. Then $\mElt{n}\hat{\te H}$ is a deviator tensor.
\end{lemma}

\begin{proof}
	The trace of $\sqrt{(\mEltl{n-1} \te C^\text{i})^{-1}} \cdot \mElt{n} \te H \cdot \sqrt{\mEltl{n-1} \te C^\text{i}}$ is given by 
	$\mElt{n}{\te H}:\left(\sqrt{(\mEltl{n-1} \te C^\text{i})^{-1}}\cdot\sqrt{\mEltl{n-1} \te C^\text{i}}\right)=\mElt{n}{\te H}:\one = 0$ since $\mElt{n}{\te H}\in\Devi$. 
\end{proof}

\begin{theorem}\label{theorem:sym_exp_map_mod}
	Let the inelastic deformation of the last time step $\mEltl{n-1} \te C^\text{i}\in\Sym\cap\SL$, i.e., $\mEltl{n-1} \te C^\text{i} = (\mEltl{n-1} \te C^\text{i})^T$ and $\det \mEltl{n-1} \te C^\text{i} = 1$, and $\mElt{n} \te H \in \Devi$, i.e., $\tr \mElt{n} \te H = 0$.
	Then the implicit exponential integrator 
	$\mElt{n}\te C{}^\text{i} = \sqrt{\mEltl{n-1} \te C^\text{i}} \cdot \exp\left(\mElt{n} \hat{\te H} \, {}^n \!\Delta t\right) \cdot \sqrt{\mEltl{n-1} \te C^\text{i}}$  with $\mElt{n}\hat{\te H}:=\sym\left(\sqrt{(\mEltl{n-1} \te C^\text{i})^{-1}} \cdot \mElt{n} \te H \cdot \sqrt{\mEltl{n-1} \te C^\text{i}}\right)$
	guarantees that the solution $\mElt{n}\te C{}^\text{i}$ is symmetric and unimodular.
\end{theorem}

\begin{proof}
	With Lemma~\ref{lemma:H_deviator} we find that $\det \mElt{n}\te C{}^\text{i} = \det\left(
	\sqrt{\mEltl{n-1} \te C^\text{i}} \cdot \exp\left(\mElt{n} \hat{\te H} \, {}^n \!\Delta t\right) \cdot \sqrt{\mEltl{n-1} \te C^\text{i}} 
	\right) = 1$. Furthermore, the symmetry of $\mEltl{n-1} \te C^\text{i}$ and $\mElt{n} \hat{\te H}$ implies that $\mElt{n}\te C{}^\text{i}$ is symmetric. Thus it holds $\mElt{n}\te C{}^\text{i} \in \Sym \cap \SL$ for the solution.
\end{proof}

In contrast to the implicit exponential integrator~\eqref{eq:exp_orig}, the modified formulation~\eqref{eq:exp_mod} automatically guarantees symmetry of the intermediate results $\mElt{n} \te C^\text{i}$ during the iterative solution.

\section{Weighting of the gate loss}
\label{app:study_gate}

\begin{figure}
	\centering
	\includegraphics{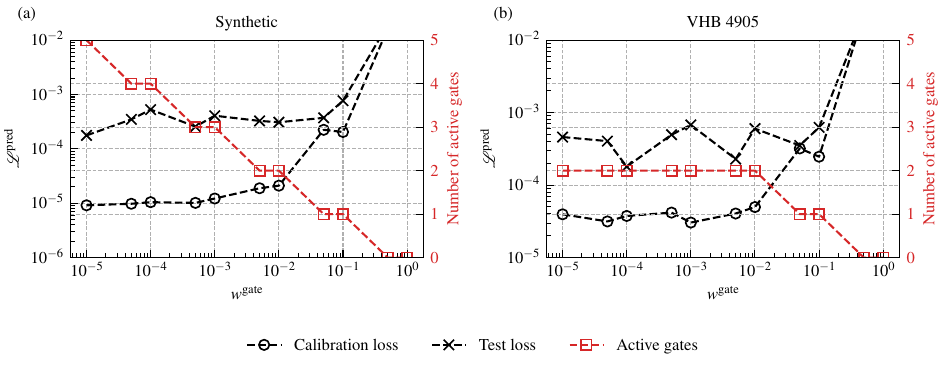}
	\caption{Variation of the weight $w^\text{gate}$ for the gate loss term  $\mathscr L^\text{gate}$: (a) synthetic data set and (b) VHB 4905. The prediction losses are the MSEs of the stresses. The results of the best run out of 5 training runs are shown.}
	\label{fig:study_sparse}
\end{figure}

In this appended section, the weight $w^\text{gate}$ for the loss term $\mathscr L^\text{gate}$ defined in Eq.~\eqref{eq:gate_loss} is varied systematically. 
The parameters for the gates and the exponent in the $p$-quasinorm are chosen to $\gamma = 1.025$, $\epsilon = 2.5$, $\delta = \num{1e-6}$, and $p=\frac{1}{4}$, respectively. The loss term for the training is given by $\mathscr L = \mathscr L^\text{pred} + w^\text{gate} \mathscr L^\text{gate}$, where the prediction loss is chosen as the MSE of the 1st Piola-Kirchhoff stresses $\te P$. The weight is varied as follows: $w^\text{gate} \in\{\num{1e-5},\num{5e-5},\ldots,\num{1e0}\}$. In all examples, the PANN models were initialized with 5 Maxwell elements, where architectures with one hidden layer have been used for all three NNs ($\psi^\text{NN}$ and $\mEl \psi^\text{NN}$ with 8 neurons in the hidden layer; $\mEl \phi^{*,\text{NN}}$ with 16 neurons in the hidden layer).

The results of the study are given in Fig.~\ref{fig:study_sparse} for the synthetic dataset and the experimental data of VHB 4905. On the left vertical axis of each subplot, the prediction loss (calibration and test) is plotted and on the right vertical axis (red) the number of active gates, i.e., gates for which the condition $g_\alpha > 0$ holds.
As can be seen, the number of active gates decreases after training as $w^\text{gate}$ increases for the synthetic data set. However, if the weight is set too high, this leads to excessive weighting of the penalty term based on the $p$-quasinorm. This initially leads to the elimination of an overly large number of Maxwell elements and, if the value is increased further, to a drastic decrease in predictive capability, as all Maxwell elements are then switched off. In the data set VHB 4905, 3 out of 5 Maxwell elements are always switched off over a wide range. Only from \num{1e-2} onwards is the penalty term weighted too heavily here.
The task is now to find a value for the weight that leads to a model with as few Maxwell elements as possible, but at the same time does not negatively affect the prediction quality. Accordingly, a value of \num{5e-3} has proven to be suitable for both cases.

\bibliographystyle{unsrtnat} 
\bibliography{finite_visco_pann.bib}

\end{document}